\documentclass[epsfig]{article}
\usepackage{amsfonts}

\usepackage{graphicx}
\usepackage{amsmath}

\input epsf.sty
\newtheorem{theorem}{Theorem}
\newtheorem{acknowledgement}[theorem]{Acknowledgement}

\input{tcilatex}
\makeatletter \@addtoreset{equation}{section}
\renewcommand{\theequation}{\thesection.\arabic{equation}}

\begin{document}

\title{\rightline{\mbox {\normalsize {Lab/UFR-HEP/0307/GNPHE/0308}}}\vspace{1cm}%
\textbf{Toric Varieties with NC Toric Actions}:\\
\textbf{\ NC Type IIA Geometry}}
\author{Mohamed Bennai$^{1,2,3,4}$\thanks{%
m.bennai@univh2m.ac.ma} and El Hassan Saidi$^{1,3}$\thanks{%
E-mail: h-saidi@fsr.ac.ma} \\
{\small 1-Lab/UFR-High Energy Physics, Physics Department, \small Faculty of Science, Rabat, Morocco%
}\textit{.}\\
{\small 2-Groupe de Physique des Hautes Energies, Fac Sciences Ben M'sik,
B.P 7955, \small Casablanca, Maroc,}\\
{\small 3- National grouping in High Energy Physics, Focal point, Faculty of Science, Rabat, Morocco,}\\
{\small 4- LPMC, Fac Sciences Ben M'sik, B.P 7955, Univ Hassan II Mohamedia,
Casablanca, Maroc.}}
\maketitle

\begin{abstract}
Extending the usual $\mathbf{C}^{\ast r}$ actions of toric manifolds by
allowing asymmetries between the various $\mathbf{C}^{\ast }$ factors, we
build a class of non commutative (NC) toric varieties $\mathcal{V}%
_{d+1}^{(nc)}$. We construct NC complex $d$ dimension Calabi-Yau manifolds
embedded in $\mathcal{V}_{d+1}^{(nc)}$ by using the algebraic geometry
method. Realizations of NC $\mathbf{C}^{\ast r}$ toric group are given in
presence and absence of quantum symmetries and for both cases of discrete or
continuous spectrums. We also derive the constraint eqs for NC Calabi-Yau
backgrounds $\mathcal{M}_{d}^{nc}$ embedded in $\mathcal{V}_{d+1}^{nc}$ and
work out their solutions. The latters depend on the Calabi-Yau condition $%
\sum_{i}q_{i}^{a}=0$, $q_{i}^{a}$ being the charges of $\mathbf{C}^{\ast r}$%
; but also on the toric data $\left\{ q_{i}^{a},\nu _{i}^{A};p_{I}^{\alpha
},\nu _{iA}^{\ast }\right\} $ of the polygons associated to $\mathcal{V}%
_{d+1}$. Moreover, we study fractional $D$ branes at singularities and show
that, due to the complete reducibility property of $\mathbf{C}^{\ast r}$
group representations, there is an infinite number of fractional $D$ branes.
We also give the generalized Berenstein and Leigh quiver diagrams for
discrete and continuous $\mathbf{C}^{\ast r}$ representation spectrums. An
illustrating example is presented. \newline
\end{abstract}

\tableofcontents

\newpage

\newpage

\section{Introduction}

Matrix model compactification of M theory on non commutative (NC) torii \cite
{a1} has opened an increasing interest in the study of non commutative
spaces, in relation with NC quantum field instantons \cite{a2}, and open
strings of the solitonic sector of type II string theories \cite{a3}-\cite
{a5}. These NC structures have found remarkable applications in various
areas of quantum physics such as in the analysis of $D(p-4)/Dp$ brane
systems ($p>3$) \cite{a6,a7} and in the study of tachyon condensation using
the GMS method \cite{a8}. However, most of NC spaces used in these studies
involve mainly NC $\mathbb{R}_{\theta }^{d}$ [9], NC $\mathbb{T}_{\theta
}^{d}$ torii \cite{a9,a10}, some cases of $\mathbb{Z}_{n}$ type orbifolds of
NC torii \cite{a11,a12} and some generalizations to non commutative higher
dimensional cycles such as the non commutative extension of Hizerbruch
complex surfaces $F_{n}$ used in \cite{a13} and some special Calabi-Yau
orbifolds.

Recently efforts have been devoted to go beyond these particular manifolds
by considering non commutative extension of complex manifolds with torsion
and too particularly NC compact Calabi-Yau manifolds $\mathcal{M}$ because
of the basic role they play in type II string compactifications and in the
geometric engineering of supersymmetric quiver gauge theories \cite{a38}.
The most studied examples are given by the class of homogeneous
hypersurfaces $\mathcal{H}_{n}$ embedded in $\mathbb{P}^{n+1}$ projective
spaces such as orbifolds of $\mathbb{K}$3 and the quintic $\mathcal{Q}$ \cite
{a14}-\cite{a18}, see also \cite{a19,a20}. The NC aspect of such varieties
is too particularly important for the stringy resolution of singularities
offering by the way an alternative method to the standard approach of
resolution by deformations of complex and Kahler structures of Calabi-Yau
manifolds. The crucial idea in this method is that in NC varieties, the
space geometry has a fine structure where the usual commutative zero
dimensional points are now represented by (matrix) operators. As a
consequence of this deformation commutative space singularities, which are
associated with the matrix identity in NC algebra, are naturally lifted;
thanks to the spectral partition property of the identity in terms of
projectors. Moreover $D$ branes wrapping NC cycles of NC Calabi-Yaus acquire
fine structures as well and then fractionate at singularities due to
complete reducibility property of matrix identity \cite{a21}-\cite{a23}. In
type II string theory, NC Calabi-Yau manifolds $\mathcal{M}^{nc}$ have
moreover a remarkable string states interpretation. The centre of $\mathcal{M%
}^{nc}$, which is just the original commutative variety $\mathcal{M}$,\ is
associated with closed string states while the NC extension is in one to one
correspondence with open string states; for details on these aspects and
related features see \cite{a39}.

Recall that the Berenstein and Leigh (BL) idea behind NC Calabi-Yau
orbifolds building consists on solving non commutativity in terms of
discrete isometries of the orbifolds. This was successfully done in \cite
{a15}; see also \cite{a24}, for the study ALE spaces and aspects of the NC
quintic; then it has been extended in \cite{a17,a18} for the building of NC
orbifolds $\mathcal{H}_{d}^{nc}$ of complex $d$-dimension homogeneous
hypersurfaces $\mathcal{H}_{d}$. In the present study, we will push this
basic idea a step further by considering a large class of CY manifolds and
introducing non commutative toric actions involving NC complex torii. In our
construction, we borrow results of BL method but think about points in NC
toric manifolds as NC torus fibers based on $\mathcal{M}$ \cite{a34}. The NC
torii involved here go beyond Connes et \textit{al} NC
\"{}%
real
\"{}%
torus solution for matrix model compactification and turns out to play a
central role in building NC toric manifolds. This way of doing can also be
thought of as a first step towards the building of non commutative extension
of supersymmetric gauged linear sigma models and their Landau Ginzburg
mirrors \cite{a40}.

To fix the ideas on BL method for NC Calabi-Yau orbifolds with discrete
torsion, let us recall one of the useful results on orbifolds of complex d
Calabi-Yau homogeneous surfaces $\mathcal{H}_{d}$. The latters are described
by the homogeneous polynomials $P_{d}\left[ z_{1},...,z_{d+2}\right]
=z_{1}^{d+2}+z_{2}^{d+2}+z_{3}^{d+2}+z_{4}^{d+2}+z_{5}^{d+2}+a_{0}\prod_{\mu
=1}^{d+2}z_{\mu }=0$ with $\mathbf{Z}_{d+2}^{d}$ discrete\ symmetries acting
as $z_{i}\longrightarrow z_{i}$.$\omega ^{q_{i}^{a}}$, where the $q_{i}^{a}$%
\ integers satisfy the\ Calabi-Yau \ condition $\sum_{i=1}^{d+2}q_{i}^{a}=0,$%
\ $a=1,...,d$. Non commutative $\mathcal{H}_{d}^{nc}$ extending complex $d$%
-dimension hypersurfaces $\mathcal{H}_{d}$\ were shown to be given, in the
coordinate patch $Z_{d+2}\sim I_{id}$, \ by the following non commutative
algebra generated by the $Z_{i}$\ operators\ satisfying,
\begin{eqnarray}
Z_{i}Z_{j} &=&\theta _{ij}\text{ }Z_{j}Z_{i};\qquad \ i,j=1,...,(d+1),
\notag \\
Z_{i}Z_{d+2} &=&\text{ }Z_{d+2}Z_{i};\qquad \ i=1,...,(d+1).
\end{eqnarray}
The $\theta _{ij}$ non commutative parameters\ are solved by discrete
torsion as $\theta _{ij}=\omega _{ij}$ $\varpi _{ji}$ with $\varpi _{kl}$
the complex conjugate of $\omega _{kl}$. The $\omega _{ij}$'s are realized
in terms of the $q_{i}^{a}$ Calabi-Yau charges and the $\mathbf{Z}_{d+2}^{d}$
discrete group elements $\omega =\exp i\frac{2\pi }{d+2}$ as $\omega
_{ij}=\exp i\left( \frac{2\pi }{d+2}m_{ab}q_{i}^{a}q_{j}^{b}\right) =\omega
^{m_{ab}q_{i}^{a}q_{j}^{b}}$ with $m_{ab}$ integers. But these relations are
nothing else than special eqs that describe a special class of complex
torii. More general construction are therefore possible.

In this paper, we aim to extend the BL non commutative geometry based on
discrete torsion \cite{a15},\cite{a17},\cite{a24},\cite{a18}, to the large
class of complex $d$ dimension Calabi-Yau manifolds $\mathcal{M}_{d}$
embedded in toric varieties $\mathcal{V}_{d+1}$ \cite{a25}-\cite{a27}. More
precisely, we will focus our attention on NC Calabi-Yau type $IIA$
geometries realized as NC hypersurfaces (NC subalgebras) in NC\ toric
manifolds though one might also do similar things for type IIB mirrors. This
analysis constitutes also the basis for field theoretic construction of NC
supersymmetric gauge sigma models and Landau Ginzburg theories. Recall that
commutative toric manifolds may, roughly speaking, be thought of as\footnote{%
Toric manifolds are generally defined by cosets $\left( \mathbf{C}%
^{k+1}-U\right) \mathbf{/C}^{\ast r}$ where \ $U\subset $ $\mathbf{C}^{k+1}$
is defined by the $\mathbf{C}^{\ast r}$\ actions and a chosen a
triangulation. In toric geometry, elements of U are defined by those subsets
of vertices, which do not lie together in a single toric cone.} $\mathcal{V}%
_{d+1}=\mathbf{C}^{k+1}\mathbf{/C}^{\ast \left( k+1-d\right) }$ and are
locally parameterized by the coordinates $\left\{ x_{i};0\leq i\leq
k\right\} $ with $\mathbf{C}^{\ast \left( k+1-d\right) }$ toric actions%
\footnote{%
The $C^{\ast r}$ toric group may, roughly speaking, be thought of as a
complexification of the $U\left( 1\right) ^{r}$\ \ gauge symmetry of two
dimensional $N=2$ supersymmetric linear sigma model. This is a continuous
abelian group \ whose representations have an elliptic sector and are
gross-modo similar to those of $U\left( 1\right) ^{r}$ representations.}.
This class of $\mathcal{M}_{d}$'s is described by hypersurfaces in $\mathcal{%
V}_{d+1}$ but the underlying polynomials $P_{d}\left[ x_{1},...,x_{k+1}%
\right] $ defining $\mathcal{M}_{d}$ are no longer homogeneous contrary to
the quintic and $\mathcal{H}_{d}$ cases. However and even though ignoring
discrete torsion,\ NC extensions of $\mathcal{M}_{d}$\ may be still obtained
by endowing the $\mathbf{C}^{\ast \left( k+1-d\right) }$ toric group with a
non commutative structure. As we will explicitly show in section 3, the non
commutative structure of resulting $\mathcal{M}_{d}^{nc}$ manifolds is
indeed induced by the symmetry of the $\mathbf{C}^{\ast r}$ toric actions
and give the basis of a more general class NC manifolds namely the NC toric
varieties. We will show moreover that solutions for non commutative geometry
have, in addition to what we were expecting, contributions coming from the
toric data $\left\{ q_{i}^{a};\nu _{i}^{A};p_{\alpha }^{I};\nu _{\alpha
A}^{\ast }\right\} $ of the polygon $\Delta \left( \mathcal{M}_{d}\right) $
of $\mathcal{M}_{d}$. The solutions for the NC constraint eqs we will
derive, follow naturally from the toric geometry identities $%
\sum_{i}q_{i}^{a}=0$ \ and $\sum_{i}q_{i}^{a}\nu _{i}^{A}=0$, with the
integers $q_{i}^{a}$\ being the charges of the $\mathbf{C}^{\ast \left(
k+1-d\right) }$ toric actions and the integers $\left( \nu _{i}^{A}\right) $%
\ the vertices of the polytope $\Delta \left( \mathcal{V}_{d+1}\right) $.
Furthermore, due to general results on representation theory of the abelian $%
\mathbf{C}^{\ast r}$ toric group, we have here also fractional $D$ branes at
the singular points of the toric actions; but with the remarkable property
that now there are infinitely many. The point is that, like for the case of
complex $N$ dimension homogeneous Calabi-Yau orbifolds with $\mathbf{Z}%
_{N+2}^{N}$ discrete groups, the NC coordinates of the $D$ branes at the
singular points are proportional to the identity operator $I_{id}$ of the
toric group representation $\mathcal{R}\left( \mathbf{C}^{\ast r}\right) $.
As the latters are completely reducible, the identity $I_{id}$\ of $\mathcal{%
R}\left( \mathbf{C}^{\ast r}\right) $\ is then decomposable into an infinite
series in the $\pi _{n}$, ($\pi \left( \alpha \right) $ for the continuous
case) projectors on the states of the representation space of $\mathcal{R}%
\left( \mathbf{C}^{\ast r}\right) $.

The organization of this paper is as follows; In section 2, we review
general aspects of Calabi-Yau manifolds $\mathcal{M}_{d}$ embedded in toric
varieties $\mathcal{V}_{d+1}$ and focus on the study of the type $IIA$
geometry of Calabi-Yau manifolds. Similar analysis may be also done for the
type $IIB$ geometry of $\mathcal{M}_{d}$; the dual of the type $IIA$.
Section 3 is mainly devoted to the study of the NC type $IIA$ geometry
extension of $\mathcal{M}_{d}$ by help of torsion of the $C^{\ast r}$ toric
actions of the toric varieties. To that purpose, we will first derive the
constraint eqs for $\mathcal{M}_{d}^{nc}$. Next we analyze different
realizations for $C^{\ast r}$ toric torsions by using quantum symmetries
generated by shift operators of the $C^{\ast r}$ toric group; but also by
introducing torsion among the $U_{a}$ generators of the $C^{\ast }$\ abelian
factors of the $C^{\ast r}$ group representations. Then we build general
solutions of the NC constraint eqs by using effectively both of these two
kinds of the $C^{\ast r}$ torsions. We end this section by giving an
illustrating example treating the NC type $IIA$ geometry. In section 4, we
study fractional branes at the singular points of the toric action and show
that, here also, we have fractional $D$ branes at singularities; but with
the basic difference that now the identity $I_{id}$ of the representation of
each $C^{\ast }$ factor of the $C^{\ast r}$ group is decomposed into an
infinite set of projectors on the representation space states. The dimension
of fractional $D$ branes are shown to be dependent on the choice of the
Calabi-Yau charges of the $C^{\ast r}$ toric actions of te toric variety $%
\mathcal{V}$. In section 5, we give our conclusion and perspectives.

\section{Calabi-Yau Hypersurfaces in Toric Varieties}

There are different ways for building complex $d$ dimension Calabi-Yau
manifolds $\QTR{sl}{M}$. A way to do is by help of 2d $\mathcal{N}=2$
supersymmetric gauged linear sigma models or again by embedding $\QTR{sl}{M}$
in a toric variety $\mathcal{V}$ [36,37]. The latter is a complex\ Kahler
manifold with some $C^{\ast r}$ toric actions generalizing the usual complex
n dimension projective spaces CP$^{n}$. The simplest Calabi-Yau example is
given by the case where $\QTR{sl}{M}$\ is described by complex $d$ dimension
hypersurface in a complex $\left( d+1\right) $ toric variety $\mathcal{V}%
_{d+1}$. To write down algebraic geometry eqs for the Calabi-Yau
hypersurfaces; one should specify a number of ingredients namely a local
holomorphic coordinates patch of the toric manifold $\mathcal{V}_{d+1}$, the
group of toric action and the toric data. To do so, one should moreover
distinguish between two kinds of geometries for the Calabi-Yau manifolds
\textsl{M}$_{d}$: (1) Type $IIA$ geometry, to which we will refer here below
to as $\mathcal{M}_{d}$, and (2) its type $IIB$ mirror geometry often
denoted as $\mathcal{W}_{d}$. The latter is\ obtained from $\mathcal{M}_{d}$
by exchanging their Kahler and complex structures following from the Hodge
identities $\ h^{1,1}\left( \mathcal{M}_{d}\right) =h^{d-1,1}\left( \mathcal{%
W}_{d}\right) $ and $h^{1,1}\left( \mathcal{W}_{d}\right) =h^{d-1,1}\left(
\mathcal{M}_{d}\right) $ \cite{a28,a29}. In this study, we will mainly focus
our attention on the type $IIA$ geometry.

Type $IIA$ geometry is constructed in terms of two dimensional $\mathcal{N}%
=2 $ supersymmetric linear sigma models as follows: First consider a
superfield system $\left\{ V_{a},X_{i}\right\} $ containing $r$ gauge $N=2$
abelian multiplets $V_{a}\left( \sigma ,\theta ,\overline{\theta }\right) $
with gauge group $U(1)^{r}$and $\left( k+1\right) $ chiral matter
superfields $X_{i}\left( \sigma ,\theta ,\overline{\theta }\right) $ of
bosonic components $x_{i}$. In addition to the usual terms namely
\begin{equation*}
\mathcal{S}\left[ V_{a},X_{i}\right] =\sum_{i}\int d^{2}\theta d^{2}%
\overline{\theta }K\left( X_{i}^{\ast }e^{2q_{i}^{a}V_{a}}X_{i}\right) ,
\end{equation*}
with $K$ being the gauge covariant Kahler superpotential and $q_{i}^{a}$ the
charges of $X_{i}$ under the $U\left( 1\right) $s, the linear sigma model
action $\mathcal{S}\left[ V_{a},X_{i}\right] $ of these fields may have $r$
Fayet Iliopoulos (FI) $D$-terms,
\begin{equation*}
\zeta _{a}\int d^{2}\sigma d^{2}\theta d^{2}\overline{\theta }V_{a}\left(
\sigma ,\theta ,\overline{\theta }\right) ,
\end{equation*}
with $\zeta _{a}$ being the FI coupling constants. The superfields action $%
\mathcal{S}\left[ V_{a},X_{i}\right] $\ may also have a holomorphic
superpotential $W(X_{0},...,X_{k})$ given by polynomials in the $X_{i}$'s,
which in the infrared limit, is known to describe a two dimensional
conformal field theory describing the string propagation on the type $IIA$
background \cite{a29}. Let us discuss a little bit this particular geometry.

\subsection{Type $IIA$ Geometry}

In the method of toric geometry, where to each complex bosonic field $x_{i}$
it is associated some toric data $\left\{ q_{i}^{a},\mathbf{\nu }%
_{i}\right\} $, or more generally by taking into account the data of the
dual geometry $\left\{ q_{i}^{a},\mathbf{\nu }_{i};p_{\alpha }^{I},\mathbf{%
\nu }_{\alpha }^{\ast }\right\} $ \cite{a19,a20}-\cite{a25}, with $\mathbf{%
\nu }_{i}$ and $\mathbf{\nu }_{\alpha }^{\ast }$\ being $\left( d+1\right) $
dimension vectors of \ $\mathbf{Z}^{d+1}$ self dual lattice, one can write
down the algebraic geometry equation of the complex $d$ Calabi-Yau $\mathcal{%
M}_{d}$ manifold. This is given by a holomorphic polynomial in the $x_{i}$'s
with some abelian complex symmetries. In the simplest situation where the
toric manifold is given by the coset $\mathcal{V}_{d+1}=\mathbf{C}%
^{k+1}/C^{\ast r}$, $d=k-r$, the complex $d$ \ dimension\ Calabi-Yau
hypersurface reads as,
\begin{equation}
P_{d}\left[ x_{0},...,x_{k}\right] =b_{0}\prod_{i=0}^{k}x_{i}+\sum_{\alpha
}b_{\alpha }\prod_{i=0}^{k}x_{i}^{n_{\alpha i}}.
\end{equation}
where the $b_{\alpha }$'s are complex structures of $\mathcal{M}_{d}$ and
where the $n_{\alpha i}$\ powers are some positive integers constrained by
the $C^{\ast r}$ invariance. Indeed, under the $C^{\ast r}$\ toric action on
the $\mathbf{C}^{k+1}$ local coordinates,
\begin{equation*}
x_{i}\longrightarrow x_{i}\lambda _{a}^{q_{i}^{a}}
\end{equation*}
with $q_{i}^{a}$\ some integers, the same as in the action $\mathcal{S}\left[
V_{a},X_{i}\right] $, invariance of $P_{d}\left[ x_{0},...,x_{k}\right] $ \
requires the $n_{\alpha i}$\ integers are such that,
\begin{equation}
\sum_{i}q_{i}^{a}n_{\alpha i}=0;\qquad \sum_{i}q_{i}^{a}=0.
\end{equation}
Eqs $\sum_{i}q_{i}^{a}=0$ \ follow from the $C^{\ast r}$ symmetry of the $%
\prod_{i=0}^{k}x_{i}$\ monomial while $\sum_{i}q_{i}^{a}n_{\alpha i}=0$ come
from invariance of $\prod_{i=0}^{k}x_{i}^{n_{\alpha i}}$ monomials.

\subsubsection{Algebraic geometry eqs}

Setting $u_{\alpha }=\prod_{i=0}^{k}x_{i}^{n_{\alpha i}}$, the above eq
(2.1) can be rewritten as $P_{d}\left[ u_{\alpha }\right] =\sum_{\alpha
}b_{\alpha }u_{\alpha }$, where the $u_{\alpha }$'s\ are the effective local
coordinates of the coset space $\mathbf{C}^{k+1}/\mathbf{C}^{\ast r}$. As
the $u_{\alpha }$\ variables are given by $u_{\alpha
}=\prod_{i=0}^{k}x_{i}^{n_{\alpha i}}$, it may happen that not all of the $%
u_{\alpha }$'s are independent variables; some of these $u_{\alpha }$, say $%
u_{\alpha _{I}}$ for $I=1,...r^{\ast }$, are expressed in terms of the other
$u_{\alpha _{I}}$ variables with $J\neq I$. In other words; one may have
relations type $\prod_{\alpha }u_{\alpha }^{p_{\alpha }^{I}}=1$, where $%
p_{\alpha }^{I}$\ are some integers. Substituting $u_{\alpha
}=\prod_{i=0}^{k}x_{i}^{n_{\alpha i}}$ back into $\prod_{\alpha }u_{\alpha
}^{p_{\alpha }^{I}}=1$, we discover extra constraint eqs on the $n_{\alpha
i} $ and $p_{\alpha }^{I}$\ integers namely,
\begin{equation}
\sum_{\alpha }p_{\alpha }^{I}n_{\alpha i}=0.
\end{equation}
In toric geometry the $n_{\alpha i}$\ integers are realized as scalar
products, $n_{\alpha i}=<\mathbf{\nu }_{i},\mathbf{\nu }_{\alpha }^{\ast
}>=\sum_{A}\nu _{i}^{A}\nu _{\alpha A}^{\ast }$, of the toric data vector
vertices $\mathbf{\nu }_{i}$ and $\mathbf{\nu }_{\alpha }^{\ast }$ of
integer entries $\nu _{i}^{A}$ and $\nu _{\alpha A}^{\ast }$ respectively.
In this representation, eqs(2.2) and (2.3) are automatically solved by
requiring the toric data of the Calabi-Yau manifold to be such that,
\begin{equation}
\sum_{i}q_{i}^{a}\mathbf{\nu }_{i}=0;\qquad \sum_{\alpha }p_{\alpha }^{I}%
\mathbf{\nu }_{\alpha }^{\ast }=0.
\end{equation}
Let us illustrate these relations for the case of the asymptotically local
Euclidean ( ALE ) space with $A_{m-1}$ singularity. This is a complex two
dimension $\mathbf{C}^{n+1}/\mathbf{C}^{\ast \left( n-1\right) }$ toric
variety with a $su(m)$ singularity $u_{1}u_{2}=u_{0}^{m}$\ at the origin.
From this relation, which can be also rewritten as $u_{0}^{-m}u_{1}u_{2}=1$,
one sees that there are three effective variables $u_{0}$, $u_{1}$ and $%
u_{2} $; but only two of them are independent. Since there is one relation,
the integer $r^{\ast }=1$ and so there is only one $p_{\alpha }^{I}$\ vector
of entries $p_{\alpha }=\left( -m,1,1\right) $ and three $\mathbf{\nu }%
_{\alpha }^{\ast }$ vectors given by $\mathbf{\nu }_{0}^{\ast }=\left(
1,0\right) $, $\mathbf{\nu }_{1}^{\ast }=\left( m,-1\right) $ and $\mathbf{%
\nu }_{2}^{\ast }=\left( 0,1\right) $. More generally, we have the following
cases: (a) In the simplest case where no relation such as $\sum_{\alpha
}p_{\alpha }^{I}\mathbf{\nu }_{\alpha }^{\ast }=0$ exist,\ that is all the $%
u_{\alpha }$'s are independent, the algebraic eq of the hypersurface $%
\mathcal{M}_{d}$ reads simply as $P_{d}\left[ u_{0},...,u_{d}\right]
=\sum_{\alpha =0}^{d}b_{\alpha }u_{\alpha }=0$ \ with $d=\left( k-r\right) $%
. (b) In the generic cases where there are for instance $r^{\ast }$\
constraint eqs of type $\sum_{\alpha }p_{\alpha }^{I}\mathbf{\nu }_{\alpha
}^{\ast }=0$, the $\left( d+r^{\ast }+1\right) $ \ complex variables $\
u_{\alpha }$ are not all of them independent and so the algebraic geometry
eqs defining the hypersurface embedded in $\mathcal{V}_{d+1}$ is
\begin{equation}
\sum_{\alpha =0}^{d+r^{\ast }}b_{\alpha }u_{\alpha }=0;\qquad \prod_{\alpha
=0}^{d+r^{\ast }}u_{\alpha }^{p_{\alpha }^{I}}=1;\qquad I=1,...,r^{\ast },
\end{equation}
where $p_{\alpha }^{I}$\ are the integers in eqs(2.4).

In the field theoretic language of the two dimensional $N=2$\ supersymmetric
linear sigma model with superfields $\left\{ V_{a},X_{i},1\leq a\leq r;0\leq
i\leq k\right\} $, the $q_{i}^{a}$ integers of the $\mathbf{C}^{\ast r}$
toric action are the $a-th$\ $\ U(1)$ charge of the $x_{i}$ fields. Recall
in passing that the $U(1)^{r}$ gauge symmetry group acts on the\ $x_{i}$
bosonic fields as $x_{i}\longrightarrow x_{i}\exp \left( iq_{i}^{a}\alpha
_{a}\right) $, with $\alpha _{a}$\ being the gauge group parameters
encountered earlier. The condition that the $N=2$ theory has an extra $R$%
-symmetry \cite{a29} is effectively given by the Calabi-Yau condition $%
\sum_{i=0}^{k}q_{i}^{a}{=0,\quad a=1,...,r}$. Moreover, for the simple case
where the $N=2$ theory has no superpotential ($W(X)=0$), the moduli space of
vacuum configurations minimizing the $D$-term scalar potential of the $N=2$
linear sigma model is generated by gauge invariant fields $u_{\alpha }$
related to the $x_{i}$'s\ as; $u_{\alpha }=\prod_{i=0}^{k}x_{i}^{<\mathbf{%
\nu }_{i},\mathbf{\nu }_{\alpha }^{\ast }>}$. This gauge invariance or
equivalently $\mathbf{C}^{\ast r}$ toric symmetry, containing as a subgroup
the $U(1)^{r}$ gauge group, of the $u_{\alpha }=\prod_{i=0}^{k}x_{i}^{<%
\mathbf{\nu }_{i},\mathbf{\nu }_{\alpha }^{\ast }>}$\ composite variables
follow from $\prod_{i=0}^{k}\lambda _{a}^{<q_{i}^{a}\mathbf{\nu }_{i},%
\mathbf{\nu }_{\alpha }^{\ast }>}=1$ which is exactly solved by the
relations $\sum_{i}q_{i}^{a}\nu _{i}^{A}=0$.

The toric manifold $\mathcal{V}_{d+1}=\mathbf{C}^{k+1}/\mathbf{C}^{\ast r}$
parameterized by the $u_{\alpha }$ variables is generically singular, but
the presence of the FI D-terms $\zeta _{a}\int d^{2}\sigma $ $D_{a}\left(
\sigma \right) $\ into the two dimensional $N=2$ action $\mathcal{S}\left[
V_{a},X_{i}\right] $ has the effect of resolving the singularity by blowing
up the manifold singularity. To fix the ideas, let us consider the type $IIA$
geometry for the complex dimension $2$ ALE space with $A_{n-1}$ singularity.
Solving the condition on the dimension, namely $k+1-r=2$ by taking $k=n$ and
$r=n-1;$ then using toric geometry method by associating to each moduli $%
x_{i}$\ the data $\left\{ q_{i}^{a};\nu _{i}^{A};p_{\alpha };\mathbf{\nu }%
_{\alpha A}^{\ast }\right\} $\ with,
\begin{eqnarray}
q_{i}^{1} &=&\left( 1,-2,1,0,...,0,0,0\right) ,\quad i=0,...,n,  \notag \\
q_{i}^{2} &=&\left( 0,1,-2,1,...,0,0,0\right) ,\quad i=0,...,n,  \notag \\
&&..., \\
q_{i}^{n-1} &=&\left( 0,0,0,0,...,1,-2,1\right) ,\quad 0=1,...,n,  \notag \\
p_{\alpha } &=&\left( -n,1,1\right)  \notag
\end{eqnarray}
while the vertices are given by,
\begin{equation}
\mathbf{\nu }_{i}=\left( 1,i\right) ,\quad \mathbf{\nu }_{\alpha }^{\ast
}=\left(
\begin{array}{cc}
1 & 0 \\
n & -1 \\
0 & 1
\end{array}
\right) ,\quad <\mathbf{\nu }_{i},\mathbf{\nu }_{\alpha }^{\ast }>=\left(
1,n-i,i\right) .
\end{equation}
These data may be expressed in an interesting compact form where one
recognizes the structure of the $su\left( n\right) $ Cartan matrix as shown
here below,
\begin{eqnarray}
q_{i}^{a} &=&\delta _{i-1}^{a}-2\delta _{i}^{a}+\delta _{i+1}^{a},\quad
a=1,...,n-1;\quad i=0,...,n,  \notag \\
\mathbf{\nu }_{i} &=&\nu _{i}^{1}\mathbf{e}_{1}+\nu _{i}^{2}\mathbf{e}_{2}=%
\mathbf{e}_{1}+i\mathbf{e}_{2};\quad i=0,...,n,  \notag \\
\mathbf{\nu }_{\alpha }^{\ast } &=&\nu _{\alpha 1}^{\ast }\mathbf{e}_{1}+\nu
_{\alpha 2}^{\ast }\mathbf{e}_{2},\quad \alpha =0,1,2, \\
\mathbf{\nu }_{0}^{\ast } &=&\mathbf{e}_{1};\quad \mathbf{\nu }_{1}^{\ast }=n%
\mathbf{e}_{1}-\mathbf{e}_{2};\quad \mathbf{\nu }_{2}^{\ast }=\mathbf{e}_{2},
\notag
\end{eqnarray}
where the two vectors $\left\{ \mathbf{e}_{1},\mathbf{e}_{2}\right\} $\ are
the generators of the $\mathbf{Z}^{2}$\ \ lattice with $\mathbf{e}_{i}\cdot
\mathbf{e}_{j}=\delta _{ij}$. Moreover, the three $u_{\alpha }$ gauge
invariant of the $\mathbf{C}^{\ast n-1}$ toric action are given by;
\begin{equation}
u_{0}=\prod_{i=0}^{n}x_{i};\quad u_{1}=\prod_{i=0}^{n}x_{i}^{n+1-i};\quad
u_{2}=\prod_{i=0}^{n}x_{i}^{i-1}
\end{equation}
From this field realization, one sees that these invariants are not all
independent since we have the constraint relation
\begin{equation*}
u_{1}u_{2}=u_{0}^{n}
\end{equation*}
showing that the complex two dimension toric manifold $\mathbf{C}^{n+1}/%
\mathbf{C}^{\ast n-1}$ we are describing has an $A_{n-1}$ singularity at the
origin $\left( u_{1,}u_{2},u_{3}\right) =\left( 0,0,0\right) $.

Using this field realization, one may also write down the one dimension
hypersurface $\mathcal{M}_{1}$ embedded in $\mathbf{C}^{n+1}/\mathbf{C}%
^{\ast n-1}$ with $A_{n-1}$ singularity. Its algebraic geometry eq $%
\sum_{\alpha =0}^{2}b_{\alpha }u_{\alpha }=0$ reads in terms of the $x_{i}$
variables as;
\begin{equation}
P_{1}\left[ x_{0},...,x_{n}\right] =b_{0}\prod_{i=0}^{n}x_{i}+b_{1}%
\prod_{i=0}^{n}x_{i}^{n+1-i}+b_{2}\prod_{i=0}^{n}x_{i}^{i-1}=0,
\end{equation}
where the $b_{\alpha }$'s are complex structure of $\mathcal{M}_{1}$.
Invariance of this polynomial under the change $x_{i}\rightarrow
x_{i}\lambda _{a}^{q_{i}^{a}}$, with $\lambda _{a}\in \mathbf{C}$, follows
from the Calabi-Yau condition $\sum_{i=0}^{n}q_{i}^{a}=0$, but also from the
following obvious identities,
\begin{equation}
\pm \sum_{i=0}^{n}iq_{i}^{a}=\pm \left( i\delta _{i-1}^{a}-2i\delta
_{i}^{a}+i\delta _{i+1}^{a}\right) =0,
\end{equation}
which are nothing but the relations $\sum_{i}q_{i}^{a}\nu _{i}^{A}=0$.

\subsubsection{$C^{\ast r}$\ Toric Symmetry}

The $\lambda _{a}$\ parameters of the $C^{\ast r}$ toric actions $%
x_{i}\longrightarrow x_{i}\lambda _{a}^{q_{i}^{a}}$ are just a kind of
complexification of the manifest and familiar $U\left( 1\right) ^{r}$ gauge
symmetry parameters $\alpha _{a}$ of \ supersymmetric gauge theories acting
on matter as $x_{i}\longrightarrow x_{i}\exp i\alpha _{a}q_{i}^{a}$. Up to
complexifying the $U\left( 1\right) ^{r}$\ symmetry; that is by replacing
the $\alpha _{a}$ real parameters by the complex ones, $\psi _{a}=\alpha
_{a}-i\rho _{a}$, $\rho _{a}\in \mathbf{R}$,\ the $U\left( 1\right) ^{r}$\
symmetry extends to the $\mathbf{C}^{\ast r}$ toric actions where now
\begin{equation*}
\lambda _{a}=\exp i\psi _{a}=\exp \left( \rho _{a}+i\alpha _{a}\right) .
\end{equation*}
As such the $U\left( 1\right) ^{r}$ gauge symmetry is recovered from the
type $IIA$ geometry by setting $\rho _{a}=0$; i.e $U\left( 1\right) ^{r}\sim
C^{\ast r}|_{\rho _{a}=0}$. Therefore the $\mathbf{C}^{\ast }$ toric
symmetries are given by the cross product of the usual $U\left( 1\right) $\
gauge symmetry with the $\mathbf{R}^{\ast }$ group acting as scaling
transformations by a real factor $\exp \rho _{a}$. Contrary to $U\left(
1\right) $,\ the $\mathbf{R}^{\ast }$\ action is not a standard symmetry in
unitary field theory; but its plays here a central role and too particularly
at the level of moduli space of type $II$ string vacuum configurations on
Calabi-Yau manifolds. For later use, it is interesting to decompose the $%
\mathbf{C}^{\ast }$ group as,
\begin{equation}
\mathbf{C}^{\ast }\mathbf{\sim R}^{\ast }\times U\left( 1\right) \sim
U\left( 1\right) \times \mathbf{R}^{\ast }.
\end{equation}
As $\mathbf{R}^{\ast }\ $and $U\left( 1\right) $ commute, $\mathbf{C}^{\ast
} $ representations, $\mathcal{R}\left( \mathbf{C}^{\ast }\right) $, are
mainly given by the tensor product of the $\mathbf{R}^{\ast }$\
representations $\mathcal{R}\left( \mathbf{R}^{\ast }\right) $ and the $%
U\left( 1\right) $\ ones $\mathcal{R}\left( U\left( 1\right) \right) $.\
Moreover like for the $U\left( 1\right) ^{r}$ symmetry, the $\mathbf{C}%
^{\ast r}$ toric group is abelian and the general properties of its
representations are grosso-modo similar to those of $U\left( 1\right) ^{r}$.
In practice, the $C^{\ast r}$ toric group may be defined as given by the set
of operators $U_{a}=\lambda _{a}^{T_{a}}$, satisfying
\begin{equation}
U_{a}U_{b}=U_{b}U_{a},\quad T_{a}T_{b}=T_{b}T_{a},
\end{equation}
and acting on the $x_{i}$'s by the following gauge transformations,
\begin{eqnarray}
q_{i}^{a}\text{ }x_{i} &=&\left[ T_{a},x_{i}\right] ,  \notag \\
U_{a} &:&x_{i}\rightarrow U_{a}x_{i}U_{a}^{-1}=x_{i}\lambda _{a}^{q_{i}^{a}}{%
.}
\end{eqnarray}
These relations show that $r$ variables $x_{i}$ among $\left\{
x_{1},...,x_{k+1}\right\} $ may be usually fixed to a constant by an
appropriate choice of the $\mathbf{C}^{\ast r}$ gauge parameters. Setting $\
U_{a}=\exp \left( i\psi _{a}T_{a}\right) ;\ \ $with $iT_{a}=t_{a}+i\tau
_{a}\quad $where $t_{a}$ and $\tau _{a}$ are hermitian operators and
substituting $\psi _{a}=\alpha _{a}-i\rho _{a}$\ \ in the expression of $%
U_{a}$, one gets
\begin{equation*}
U_{a}=\exp i\left( \rho _{a}K_{a}+i\alpha _{a}Q_{a}\right)
\end{equation*}
with $K_{a}=\frac{\alpha _{a}}{\rho _{a}}t_{a}+\tau _{a}$\ \ and $Q_{a}=\tau
_{a}-\frac{\rho _{a}}{\alpha _{a}}t_{a}$\ \ generating $\mathbf{R}^{\ast r}$
and $U\left( 1\right) ^{r}$ respectively. Since here $\mathbf{R}^{\ast
}\times U\left( 1\right) \sim U\left( 1\right) \times \mathbf{R}^{\ast }$,
we have also
\begin{equation}
\left[ K_{a},K_{b}\right] =0;\quad \left[ Q_{a},Q_{b}\right] =0;\quad \left[
K_{a},Q_{b}\right] =0.
\end{equation}
Dilatations generated by $K_{a}$ and phase transformations generated by $%
Q_{b}$\ commute; they may be diagonalized simultaneously on a basis of the
representation vector space of $\mathbf{C}^{\ast r}$.

\subsection{More on Type $IIA$ Geometry}

From the previous analysis, we have learnt that generic forms of the
algebraic equations of the type $IIA$ geometry of complex $d-$dimension
Calabi-Yau manifolds are generally given by the following polynomials,
\begin{equation}
P_{d}\left[ x_{0},...,x_{k}\right] =b_{0}\prod_{i}x_{i}+\sum_{\alpha
}b_{\alpha }\prod_{i}x_{i}^{n_{\alpha i}}
\end{equation}
where $x_{i}$ are holomorphic homogeneous variables satisfying $%
x_{i}\rightarrow x_{i}\lambda _{a}^{q_{i}^{a}}$, the $b_{\alpha }$'s are
complex structures of the type $IIA$ geometry $\mathcal{M}_{d}$ of the
Calabi-Yau manifold and their number is a priori given by $h^{d-1,1}\left(
\mathcal{M}_{d}\right) $. One of the basic property of this algebraic
geometry eq is that its invariance under the change $x_{i}\rightarrow
x_{i}\lambda _{a}^{q_{i}^{a}}$, follows from the Calabi-Yau condition $%
\sum_{i}q_{i}^{a}=0$, but also due to the special relations \ $%
\sum_{i}q_{i}^{a}n_{\alpha i}=0,\quad a=1,...,r;\quad \alpha =1,....$, which
have no analogue in the case of isometries of discrete torsions used in \cite
{a15,a17}. Setting $N_{\alpha i}^{a}=q_{i}^{a}n_{\alpha i}$,\ this relation
may be also rewritten as
\begin{equation}
\sum_{i=0}^{k}N_{\alpha i}^{a}=0,\quad a=1,...,r;\quad \alpha =1,....
\end{equation}
In addition to the $N_{\alpha i}^{a}$ integers, $N_{\alpha i}^{a}=q_{i}^{a}%
\mathbf{\nu }_{i}\cdot \mathbf{\nu }_{\alpha }^{\ast }$, one also define,
out of the toric data $\left\{ q_{i}^{a};\nu _{i}^{A};p_{\alpha }^{I};\nu
_{\alpha A}^{\ast }\right\} $, others sets of integers with some specific
properties\ useful when we study the NC toric manifolds. May be the most
natural ones are those given by $N_{ij}^{ab}=q_{i}^{a}q_{j}^{b}\mathbf{\nu }%
_{i}\cdot \mathbf{\nu }_{j}$ \ satisfying
\begin{equation}
\sum_{i=0}^{k}N_{ij}^{ab}=\sum_{j=0}^{k}N_{ij}^{ab}=0.
\end{equation}
The $N_{ij}^{ab}$\ object is in fact a particular tensor of a more general
one namely,
\begin{equation*}
N_{ij}^{abAB}=q_{i}^{a}q_{j}^{b}\nu _{i}^{A}\nu _{j}^{B},
\end{equation*}
the latters still obey the previous relations and by summing on the capital
indices, one gets $N_{ij}^{ab}$. Moreover setting $A=B=1$ and taking into
account $\nu _{i}^{1}=1$, we get the first useful set of integers,
\begin{equation}
L_{ij}^{\left( 1\right) }=m_{\left[ ab\right] }q_{i}^{a}q_{j}^{b}=m_{ab}%
\text{ }q_{[i}^{a}q_{j]}^{b},
\end{equation}
where $m_{\left[ ab\right] }=\left( m_{ab}-m_{ba}\right) $ is an
antisymmetric $r\times r$ matrix with $\frac{r\left( r-1\right) }{2}$
entries. The $L_{ij}^{\left( 1\right) }$ is also a $r\times r$ antisymmetric
matrix with $\frac{r\left( r-1\right) }{2}$ entries; it satisfies the
identities
\begin{equation*}
\sum_{i}L_{ij}^{\left( 1\right) }=\sum_{j}L_{ij}^{\left( 1\right) }=0
\end{equation*}
\ inherited from the Calabi-Yau condition $\sum_{i}q_{i}^{a}=0$. Using the
toric data vertices $\nu _{i}^{A}$ of $\mathcal{V}$, one may also define an
other antisymmetric tensor $L_{ij}^{\left( 2\right) }$ satisfying as well
the identity $\sum_{i}L_{ij}^{\left( 2\right) }=\sum_{j}L_{ij}^{\left(
2\right) }=0$ inherited from the condition $\sum_{i=0}^{k}q_{i}^{a}\nu
_{i}^{A}=0$. This is given by further contracting the $A$ and $B$ indices of
$m_{ab}N_{ij}^{abAB}$by a tensor metric $m_{AB}$ as shown here below,
\begin{equation}
L_{ij}^{\left( 2\right) }=Q_{i}^{aA}Q_{j}^{bB}\left( m_{\left[ ab\right]
}m_{\left( AB\right) }+m_{\left( ab\right) }m_{\left[ AB\right] }\right) .
\end{equation}
Here $m_{ab}$ is the matrix appearing in eqs(2.19) and $m_{AB}$ is a priori
a $\left( d+1\right) \times \left( d+1\right) $ matrix. Later on, when we
study the NC Calabi-Yau manifolds, we will consider only the $m_{\left(
AB\right) }$ symmetric part and \ taking it as $m_{AB}=\epsilon _{A}\epsilon
_{B}$, where$\ \epsilon _{A}$ are numbers. Therefore $L_{ij}^{\left(
2\right) }$ have now $\left( d+1\right) $ degrees of freedom in addition to
those coming from $m_{ab}$ and which were already counted. There is moreover
a third term $L_{ij}^{\left( 3\right) }$ with the same properties as $%
L_{ij}^{\left( 1\right) }$\ and $L_{ij}^{\left( 2\right) }$. This term
involves quadratic terms in $N_{\alpha i}^{a}\ $\ and reads as,
\begin{equation}
L_{ij}^{\left( 3\right) }=N_{\alpha i}^{a}N_{\beta j}^{b}\left( m_{\left[ ab%
\right] }m^{\left( \alpha \beta \right) }+m_{\left( ab\right) }m^{\left[
\alpha \beta \right] }\right) .
\end{equation}
It satisfies the antisymmetry property $L_{ij}^{\left( 3\right)
}=-L_{ji}^{\left( 3\right) }$ and the generalized Calabi-Yau condition $%
\sum_{i}L_{ij}^{\left( 3\right) }=0$ following from $\sum_{i}N_{\alpha
i}^{a}=0$. Like for $L_{ij}^{\left( 2\right) }$, the $m^{\alpha \beta }$
matrix will be taken as $m^{\alpha \beta }=\epsilon ^{\alpha }\epsilon
^{\beta }$. Note that the sum $L_{ij}=L_{ij}^{\left( 1\right)
}+L_{ij}^{\left( 2\right) }+L_{ij}^{\left( 3\right) }$ is also an
antisymmetric matrix and has the remarkable form $%
L_{ij}=u_{[i}^{a}v_{j]}^{a} $\ verifying the the generalized Calabi-Yau
identity $\sum_{i}u_{i}^{a}=0$, \ $\sum_{i}v_{i}^{a}=0$ and so;
\begin{equation}
\sum_{i}L_{ij}=0.
\end{equation}
In the next section, we will give more details about these special features
and the way they enter in the building of the NC type $IIA$ geometry. We
will mainly focus our attention on Calabi-Yau hypersurfaces $\mathcal{M}_{d}$
embedded in $\mathcal{V}_{d+1}$; but a similar analysis may be also done for
the toric variety itself.

\section{NC Type $IIA$ Geometry}

From the algebraic geometry point of view, the NC extension $\mathcal{M}%
_{d}^{nc}$ of the Calabi-Yau manifold\ $\mathcal{M}_{d}$, embedded in $%
\mathcal{V}_{d+1}$,\ is covered by a finite set of holomorphic operator
coordinate patches $\mathcal{U}_{(\alpha )}=\{Z_{i}^{(\alpha )};1\leq i\leq
k\;\alpha =1,2,\ldots \}$ \ and holomorphic transition functions mapping \ $%
\mathcal{U}_{(\alpha )}$\ \ to \ $\mathcal{U}_{(\beta )}$;
\begin{equation*}
\phi _{(\alpha ,\beta )}:\ \mathcal{U}_{(\alpha )}\ \ \rightarrow \ \mathcal{%
U}_{(\beta )}.
\end{equation*}
This is equivalent to say that $\mathcal{M}_{d}^{nc}$ is covered by a
collection of non commutative local algebras $\mathcal{M}_{d}^{nc}\mathcal{{%
_{(\alpha )}}}$ generated by the analytic coordinate of the $\mathcal{U}%
_{(\alpha )}$ patches of $\mathcal{M}_{d}^{nc}$, together with analytic maps
$\phi _{(\alpha ,\beta )}$ on how to glue $\mathcal{M}_{d(\alpha )}^{nc}$
and $\mathcal{M}_{d}^{nc}\mathcal{{_{(\beta )}}}$. The $\mathcal{M}_{d}^{nc}%
\mathcal{{_{(\alpha )}}}$ algebras have centers $\mathcal{Z}_{\left( \alpha
\right) }=\mathcal{Z}\left( \mathcal{M}_{d(\alpha )}^{nc}\right) $; \ when
glued together give precisely the commutative manifold $\mathcal{M}_{d}$. In
this way a singularity of $\mathcal{M}_{d}\mathcal{\simeq {Z}}\left(
\mathcal{M}_{d}^{nc}\right) $ can be made smooth in the non commutative
space $\mathcal{M}_{d}^{nc}$ \cite{a24,a18}. This idea was successfully used
to build NC ALE spaces and some realizations of Calabi-Yau threefolds such
as the quintic threefolds $\mathcal{Q}$ described by the homogeneous
hypersurface \cite{a15,a17}
\begin{equation*}
P_{3}\left[ z_{1},...,z_{5}\right]
=z_{1}^{5}+z_{2}^{5}+z_{3}^{5}+z_{4}^{5}+z_{5}^{5}+b_{0}\prod_{i=1}^{5}z_{i}.
\end{equation*}
In this context, it was shown that the non commutative quintic $\mathcal{Q}%
^{nc}$ extending\ $\mathcal{Q}$,\ when expressed in the coordinate patch $%
Z_{5}=I_{id}$, \ is given by the following special algebra,
\begin{eqnarray}
Z_{1}Z_{2} &=&\alpha \ Z_{2}Z_{1},\qquad Z_{3}Z_{4}=\beta \gamma \
Z_{4}Z_{3},\qquad \qquad (a)  \notag \\
Z_{1}Z_{4} &=&\beta ^{-1}\ Z_{4}Z_{1},\qquad Z_{2}Z_{3}=\alpha \gamma \
Z_{3}Z_{2},\qquad \qquad (b)  \notag \\
Z_{2}Z_{4} &=&\gamma ^{-1}\ Z_{4}Z_{2},\qquad Z_{1}Z_{3}=\alpha ^{-1}\beta \
Z_{3}Z_{1},\qquad \qquad (c) \\
Z_{i}Z_{5} &=&\ Z_{5}Z_{i},\qquad i=1,2,3,4;  \notag
\end{eqnarray}
where $\alpha ,\beta $\ and\ $\gamma $\ are fifth roots of the unity, the
parameters of the $\mathbf{Z}_{5}^{3}$\ discrete group and where the $Z_{i}$%
\ 's are the generators of $\mathcal{Q}^{nc}$. One of the main features of
this non commutative algebra is that its centre $\mathcal{Z(Q}^{nc})$
coincides exactly with $\mathcal{Q}$, the commutative quintic threefolds. In
\cite{a15} , a special solution for this algebra using $5\times 5$\ matrices
has been obtained and in \cite{a17} a class of solutions for eqs(1.1)
depending on the Calabi-Yau charges of the quintic threefold has been worked
out and partial results regarding higher dimensional Calabi-Yau
hypersurfaces were given. A more involved analysis addressing the question
of the explicit dependence into the discrete torsion of the orbifold group,
the varieties of the fractional $D$ branes at singularities and more
generally fractional branes on NC toric manifolds have discussed recently in
\cite{a18}.

One of the key points in the building of $\mathcal{Q}^{nc}$ is the use of
discrete torsions of the symmetry $\ z_{i}\rightarrow z_{i}$ $\omega
^{q_{i}^{a}}$, $\omega ^{5}=1$, of the hypersurface $%
z_{1}^{5}+z_{2}^{5}+z_{3}^{5}+z_{4}^{5}+z_{5}^{5}+b_{0}%
\prod_{i=1}^{5}z_{i}=0 $. By working in the coordinate patch $z_{5}=1$, then
associating to each $z_{i}$\ \ variable, an operator $Z_{i}$ with $Z_{5}\sim
I_{id}$ and finally requiring that the $Z_{i}^{5}$ and $\prod_{i=1}^{5}Z_{i}$
monomials have to be in the centre of $\mathcal{Q}^{nc}$, one gets
constraint eqs when solved give the explicit expression of the $Z_{i}$\
operators in terms of the generators of the orbifold group symmetry of the
quintic.

\subsection{NC Toric Varieties}

To build the non commutative type $IIA$ geometry extending the manifold $%
\mathcal{M}_{d}$, we will more a less adopt the same method introduced in
\cite{a15,a17}. We start from the complex hypersurface $P_{d}\left[
x_{0},...,x_{k}\right] $ eq(2.16), with $(x_{0},...,x_{k})$ being the
homogeneous variables of $\mathbf{C}^{k+1}\mathbf{/C}^{\ast r}$. This
polynomial has a set of continuous isometries acting on the homogeneous
coordinates $x_{i}$ as $x_{i}\rightarrow x_{i}\lambda _{a}^{q_{i}^{a}}$. The
main difference between these $\mathbf{C}^{\ast r}$ actions and the discrete
symmetry $z_{i}\rightarrow z_{i}$ $\omega ^{q_{i}^{a}}$ used in building of $%
\mathcal{Q}^{nc}$ is that its algebraic geometry eq is given by a
homogeneous polynomial constraining $\omega $ \ to take a finite set of
discrete values,( $\omega ^{5}=1$ for \ $\mathcal{Q}$). As the polynomial
eq(2.16) describing the type $IIA$ geometry is not a homogeneous polynomial,
the $\lambda _{a}$'s are arbitrary non zero $C$-numbers subject to no
condition and so one expects emergence of a rich NC structure.

\subsubsection{Constraint Eqs}

Extending naively the algebraic geometry method used for $\mathcal{Q}^{nc}$
to our present case by associating to each $x_{i}$\ variable the operator $%
Z_{i}$, then taking $q_{k}^{a}=0$ and working in the coordinate patch $%
x_{k}=1$, or equivalently\ $Z_{k}=I_{id}$, the NC type $IIA$ geometry $%
\mathcal{M}_{d}^{nc}$ may be defined as,
\begin{eqnarray}
Z_{i}Z_{j} &=&\theta _{ij}Z_{j}Z_{i},\quad i,j=0,...,k  \notag \\
Z_{k}Z_{i} &=&Z_{k}Z_{i}.
\end{eqnarray}
Since $\mathcal{M}_{d}$\ should be in the centre of $\mathcal{M}_{d}^{nc}$,
it follows that the $Z_{i}$ generators should satisfy the constraint eqs,
\begin{equation}
\left[ Z_{i},\prod_{j=0}^{k}Z_{j}^{n_{\alpha j}}\right] =0,
\end{equation}
The non zero $\theta _{ij}$ parameters carrying the non commutativity
structure of $\mathcal{M}_{d}^{nc}$\ are then constrained as,
\begin{equation}
\prod_{j=0}^{k}\theta _{ij}=1,\quad \forall i,\qquad \theta _{ij}\theta
_{ji}=1
\end{equation}
Actually these relations constitute the defining conditions of non
commutative type $IIA$ geometry $\mathcal{M}_{d}^{nc}$. While the constraint
relation $\theta _{ij}\theta _{ji}=1$\ shows that $\theta _{ij}=\theta
_{ji}^{-1}$, the solution of the constraint eqs $\prod_{j=0}^{k}\theta
_{ij}=1$ are not trivial and should be expressed in terms of the toric data $%
\left\{ q_{i}^{a};\nu _{i}^{A};p_{\alpha }^{I};\nu _{\alpha A}^{\ast
}\right\} $ of the toric variety.

\subsubsection{Comments}

We give two comments; the first one concerns the above construction which
may be given a deeper explanation. As the Calabi-Yau manifold $\mathcal{M}%
_{d}$ is realized as a hypersurface in $\mathcal{V}_{d+1}$; it is natural to
demand that $\mathcal{M}_{d}^{nc}$ to be also given by a non commutative
subalgebra of a NC toric variety $\mathcal{V}_{d+1}^{nc}$ with $\mathbf{C}%
^{\ast r}$\ toric actions with \textit{torsions \ }$\mathbf{\tau }_{ab}$.
The non commutative structure of $\mathcal{V}_{d+1}^{nc}$ is induced by
these torsions and the original commutative toric manifold $\mathcal{V}%
_{d+1} $ is in its centre; i.e $\mathcal{V}_{d+1}=\mathcal{Z}\left( \mathcal{%
V}_{d+1}^{nc}\right) $. Using the previous correspondence $%
x_{i}\longrightarrow Z_{i}$, the NC toric variety may, locally, be defined
by the NC algebra
\begin{equation}
Z_{i}Z_{j}=\theta _{ij}Z_{j}Z_{i},
\end{equation}
together with
\begin{eqnarray}
U_{a}Z_{i} &=&\mu _{ai}\text{ }Z_{i}U_{a},  \notag \\
U_{a}U_{b} &=&\vartheta _{ab}\text{ }U_{b}U_{a},
\end{eqnarray}
where $\mu _{ai}$ and $\vartheta _{ab}$\ are non zero complex numbers. The
NC extension $\mathcal{V}_{d+1}^{nc}$\ of the toric variety $\mathbf{C}^{k+1}%
\mathbf{/C}^{\ast r}$ may be then thought of as given by $\mathbf{C}_{%
\mathbf{\theta }}^{k+1}\mathbf{/C}_{\mathbf{\tau }}^{\ast r}$, the
deformation parameters of the NC $\mathbf{C}_{\mathbf{\theta }}^{k+1}$\
space and the NC $\mathbf{C}_{\mathbf{\tau }}^{\ast r}$ toric group are
respectively $\theta =\left( \theta _{ij}\right) $ \ and \ $\mathbf{\tau }%
_{ab}\sim \log \vartheta _{ab}$. The centre $\mathcal{V}_{d+1}=\mathcal{Z}%
\left( \mathcal{V}_{d+1}^{nc}\right) $ of the NC toric variety is generated
by the $\mathbf{C}_{\tau }^{\ast r}$\ invariants $u_{\alpha }$ satisfying
\begin{equation}
\left[ Z_{i},u_{\alpha }\right] =\left[ U_{a},u_{\alpha }\right] =0;\quad %
\left[ u_{\alpha },u_{\beta }\right] =0,
\end{equation}
which coincide exactly with eqs(3.3) defining $\mathcal{V}_{d+1}$.

To summarize $\mathcal{M}_{d}^{nc}$ is a subalgebra of $\mathcal{V}%
_{d+1}^{nc}$ and $\mathcal{M}_{d}$\ is contained in the centre $\mathcal{Z}%
\left( \mathcal{V}_{d+1}^{nc}\right) $ of the NC toric variety. $\mathcal{V}%
_{d+1}^{nc}$\ and $\mathcal{M}_{d}^{nc}$ are generated by the $Z_{i}$\
operators while $\mathcal{V}_{d+1}$ and $\mathcal{M}_{d}$ \ are generated by
the $\mathbf{C}_{\tau }^{\ast r}$ invariants; the first by the equation $%
\prod_{\alpha }u_{\alpha }^{p_{\alpha }^{I}}=1$, with $u_{\alpha
}=\prod_{i}Z_{i}^{n_{\alpha i}}$, and the second by its hypersurface $%
\sum_{\alpha }u_{\alpha }=0$. Thus we have the following picture,
\begin{eqnarray}
\mathcal{M}_{d}^{nc} &\subset &\mathcal{V}_{d+1}^{nc}  \notag \\
\mathcal{M}_{d} &=&\mathcal{Z}\left( \mathcal{M}_{d}^{nc}\right) \subset
\mathcal{Z}\left( \mathcal{V}_{d+1}^{nc}\right) .
\end{eqnarray}
The second comment we want to give deals with extra discrete symmetries of
the commutative toric variety $\mathcal{V}_{d+1}$ described by the complex
eq $\prod_{\alpha }u_{\alpha }^{p_{\alpha }^{I}}=1$. If we denote by $\omega
_{\alpha }$ with $\omega _{\alpha }^{m_{\alpha }}=1$, $m_{\alpha }$\
integers\ generating a discrete group $\Gamma $, and performing the change \
$u_{\alpha }\longrightarrow \omega _{\alpha }^{n_{\alpha }}$ $u_{\alpha }$,
then invariance of \ $\prod_{\alpha }u_{\alpha }^{p_{\alpha }^{I}}=1$\ \
under $\Gamma $ requires the following relation to hold,
\begin{equation}
\sum_{\alpha }n_{\alpha }\text{ }p_{\alpha }^{I}=0.
\end{equation}
The simplest example is given by the ALE space with a $su\left( N\right) $
singularity described by eq $u_{1}u_{2}u_{0}^{-N}=1$. In this case the
discrete group is $Z_{N}$, that is $\omega _{\alpha }=\exp i\frac{2\pi }{N}$%
\ and so the $n_{\alpha }$ numbers of eq(3.9) are constrained as,
\begin{equation}
n_{0}N-n_{1}-n_{2}=0\quad \func{mod}N,
\end{equation}
which is naturally solved by the special choice $n_{1}=1$, $n_{2}=-1$ and $%
n_{0}\in \mathbf{Z}$. \ It is the torsion of this kind of discrete
symmetries that has been considered in [24] \ for building NC ALE spaces.
For our concerns, we expect that this relation and above all the relation $%
\sum_{\alpha }p_{\alpha }^{I}\mathbf{\nu }_{\alpha }^{\ast }=0$ \ may play
an important role in the building of NC type $IIB$ geometry using mirror
symmetry.

\subsection{Solving the Constraint Eqs}

First of all note that since the $\theta _{ij}$\ 's\ are non zero
parameters, one may set
\begin{equation}
\theta _{ij}=\prod_{a,b=1}^{r}\eta _{ab}^{J_{ij}^{ab}};\quad \eta _{ab}=\exp
\left( \beta _{a}\beta _{b}\right) ;\quad \beta _{a}\in \mathbf{C,}
\end{equation}
and solve the constraint eqs(3.4) by introducing torsions for the $\mathbf{C}%
^{\ast r}$ toric actions. Putting this parameterisation back into eqs(3.4),
one gets the following constraint on the $J_{ij}^{ab}$'s,
\begin{equation}
\sum_{i=0}^{k}J_{ij}^{ab}=0;\quad J_{ij}^{ab}=-J_{ji}^{ab}.
\end{equation}
Moreover as we are looking for a non commutative structure induced from
torsions of the $\mathbf{C}^{\ast r}$ toric actions and solutions to the $%
Z_{i}$\ operators as monomials in terms of the $\mathbf{C}^{\ast r}$\ group
representation generators, let us start by studying $\mathbf{C}^{\ast r}$\
toric groups with torsions; then turn to build the solutions for eqs(3.4).
Lessons from representations of NC torii and orbifolds with discrete torsion
teach us that we should distinguish to main cases depending on whether
quantum symmetries are taken into account or not.

\subsubsection{NC $C^{\ast r}$ Toric Actions}

The $\mathbf{C}^{\ast r}$\ toric group as used in toric geometry is a
complex abelian group which reduces to $U\left( 1\right) ^{r}$ once the
group parameters $\lambda _{a}=\exp i\psi _{a}$ are chosen as $\left|
\lambda _{a}\right| =1$, that is $\lambda _{a}=\exp i\alpha _{a}$, the $%
\alpha _{a}$'s real numbers and $\rho _{a}=0$. The\ $\mathbf{C}^{\ast r}$\
toric group reduces further to a discrete symmetry $\mathbf{Z}_{N_{1}}\times
...\times \mathbf{Z}_{N_{r}}$ if all $\alpha _{a}$'s\ are chosen as $\alpha
_{a}=\frac{2\pi }{N_{a}}$, with $N_{a}$ integers. Therefore, we expect that
several features regarding non commutative extensions of the $C^{\ast r}$\
toric actions\ to be generalization of known results of NC torii and
orbifolds with discrete symmetries. One of the remarkable features concerns
the analogue of the quantum symmetry which we want to consider now.

\paragraph{\textbf{1. Quantum \ toric symmetry }}

\qquad To better illustrate the introduction of torsion via quantum toric
symmetries, we consider the simple case $r=1$ or again the case of a $%
\mathbf{C}^{\ast }$\ factor of the $\mathbf{C}^{\ast r}$ toric group. Since $%
C^{\ast }$ is an abelian continuous group and its representations have very
special features, we have to distinguish the usual cases; (a) the discrete
infinite dimensional spectrum representation and (b) the continuous one.
Both of these realizations are important for the present study and should be
thought of as extensions of the irrational representations of NC real torii
\cite{a10,a12}.

\textbf{a) Discrete Spectrum}\newline
Let $\mathcal{R}_{dis}\left( C^{\ast }\right) =\left\{ U=\exp i\psi
T\right\} $ denote a representation of $C^{\ast }$ on an infinite
dimensional space $\mathbf{E}_{dis}$ with a discrete spectrum generated by
the orthonormal vector basis \footnote{%
We will use the convention notation $n\equiv n_{1}+in_{2}\in \mathbf{%
Z+iZ\sim }\left( n_{1},n_{2}\right) \in \mathbf{Z}^{2}$; $|n>$ should be
then thought of as $|n_{1}>\otimes |n_{2}>$\ .}
\begin{equation*}
\left\{ |n>;n=\left( n_{1},n_{2}\right) \in \mathbf{Z\times Z\sim Z}%
^{2}\right\}
\end{equation*}
. Here $\psi $ is a complex parameter, $\psi \in \mathbf{C}$; and $T$ is the
complex generator of $C^{\ast }$; the $\psi T$ combination may be split as $%
\psi T=\rho K+i\alpha Q$, where $K$\ is the generator of dilatations and $Q$
is the generator of phases. For $\psi =\alpha $ real ( $\rho =0$\ ), the
group representation $\mathcal{R}_{dis}\left( C^{\ast }\right) $ reduces to $%
\mathcal{R}_{dis}\left( U\left( 1\right) \right) =\left\{ U=\exp i\alpha
Q\right\} $; the usual $U\left( 1\right) $ gauge group representation for
the $N=2$ supersymmetric linear sigma model.

The generator $T$ of $\mathcal{R}_{dis}\left( C^{\ast }\right) $\ acts
diagonally on the vector basis $\left\{ |n>\right\} $; i.e \ $T|n>=n|n>$\
and so the representation group element $U$ acts as $U|n>=\left( \exp i\psi
n\right) |n>$. The representation elements are also diagonal and read as $%
U=\sum_{n}$ $\chi _{n}\left( \psi \right) $ $\pi _{n}$ where the $\chi
_{n}\left( \psi \right) $\ characters are given by $\exp \left( i\psi
n\right) $. The $\pi _{n}$'s, $\pi _{n}=|n><n|$, are the\ projectors on the
states $|n>$ of the $\mathbf{E}_{dis}$ representation space and may also
expressed as $\pi _{n}=\sum_{n}\exp \left( -i\psi n\right) U_{n}$. Since the
representation $\mathcal{R}_{dis}\left( C^{\ast }\right) $ is completely
reducible, its $I_{id}$ identity operator may be decomposed in a series of $%
\pi _{n}$ as,
\begin{equation}
I_{id}=\sum_{n}\pi _{n}.
\end{equation}
Such a relation should be compared with analogous ones for $U\left( 1\right)
$ and more particularly abelian discrete groups such as the $\mathbf{Z}_{N}$
symmetries appearing in the well known $\mathbf{C}^{2}/\mathbf{Z}_{N}$
orbifolds.

Like for $U\left( 1\right) $ and the $\mathbf{Z}_{N}$ \ discrete symmetries,
we have here also a complex shift operator $V_{\tau _{\left( 1,1\right)
}}\equiv V_{\tau }$ of the $C^{\ast }$\ group\ acting on $\left\{ |n>;n\in
\mathbf{Z+iZ}\right\} $ as an automorphism exchanging the $\mathbf{C}^{\ast
} $\ characters $\chi _{n}\left( \psi \right) $. This translation operator
which \ operates as $V_{\tau }|n>=|n+\tau >$; with $\tau _{\left( 1,1\right)
}=1+i$, may also be defined by help of the $\mathbf{a}_{\left(
n_{1},n_{2}\right) }^{+}=|\left( n_{1}+1\right) +in_{2}><n_{1}+in_{2}|$ \
and \ $\mathbf{b}_{\left( n_{1},n_{2}\right) }^{+}=|n_{1}+i\left(
n_{2}+1\right) ><n_{1}+in_{2}|$\ \ step operators\ as,
\begin{eqnarray}
V_{\left( 1,1\right) } &=&V_{1}V_{i}  \notag \\
V_{1} &=&\sum_{n_{1},n_{2}\in \mathbf{Z}}\mathbf{a}_{\left(
n_{1},n_{2}\right) }^{+};\quad V_{i}=\sum_{n_{1},n_{2}\in \mathbf{Z}}\mathbf{%
b}_{\left( n_{1},n_{2}\right) }^{+}.
\end{eqnarray}
Due to the remarkable property $\mathbf{a}_{\left( n_{1},n_{2}\right)
}^{+}\pi _{\left( n_{1},n_{2}\right) }=\pi _{\left( n_{1}+1,n_{2}\right) }%
\mathbf{a}_{\left( n_{1},n_{2}\right) }^{+}$ , $\mathbf{b}_{\left(
n_{1},n_{2}\right) }^{+}\pi _{\left( n_{1},n_{2}\right) }=\pi _{\left(
n_{1},n_{2}+1\right) }\mathbf{b}_{\left( n_{1},n_{2}\right) }^{+}$\ and \ $%
\mathbf{b}_{\left( n_{1},n_{2}\right) }^{+}\mathbf{a}_{\left(
n_{1},n_{2}\right) }^{+}\pi _{\left( n_{1},n_{2}\right) }=\pi _{\left(
n_{1}+1,n_{2}+1\right) }\mathbf{b}_{\left( n_{1},n_{2}\right) }^{+}\mathbf{a}%
_{\left( n_{1},n_{2}\right) }^{+}$\ , it follows that the operators $U$ and $%
V$ satisfy the following non commutative algebra,
\begin{equation}
UV=e^{-i\psi \tau }VU,
\end{equation}
describing a complex version of the CDS torus \cite{a1};\ to which we shall
refer to as the non commutative complex two torus. Since $\psi $\ is an
arbitrary complex parameter, eqs(3.15) define an irrational discrete
representation of the NC complex two torus. This representation satisfy the
following natural relation, useful when we discuss the solution of the
constraint eqs(3.4).
\begin{equation}
U^{k}V^{l}=\left( e^{-i\psi \tau }\right) ^{kl}V^{l}U^{k}.
\end{equation}
Before going ahead, let us make three remarks regarding the representations
of the $\mathbf{C}^{\ast }$ toric group. The first remark is that as far as
the factor $C^{\ast }\sim \mathbf{R}^{\ast }\times U\left( 1\right) $ is
concerned, one should distinguish four sectors for $\mathcal{R}\left(
C^{\ast }\right) $ according to the spectrums of its subgroup
representations $\mathcal{R}\left( \mathbf{R}^{\ast }\right) $ and $\mathcal{%
R}\left( U\left( 1\right) \right) $:

\textit{(\textbf{i}) Discrete-discrete sector \qquad }denoted as $\mathcal{R}%
_{\left( dis,dis\right) }\left( \mathbf{C}^{\ast }\right) $, this sector has
discrete spectrums for both of the two subgroup representations $\mathcal{R}%
_{dis}\left( \mathbf{R}^{\ast }\right) $ and $\mathcal{R}_{dis}\left(
U\left( 1\right) \right) $ of $\mathcal{R}_{\left( dis,dis\right) }\left(
C^{\ast }\right) ;$ that is $\mathcal{R}_{dis}\left( \mathbf{R}^{\ast
}\right) \sim \mathbf{Z}$\ \ and $\mathcal{R}_{dis}\left( U\left( 1\right)
\right) \sim \mathbf{Z}$\ and so,
\begin{equation}
\mathcal{R}_{\left( dis,dis\right) }\left( C^{\ast }\right) \sim \mathbf{%
Z\times Z}
\end{equation}
This is the case we have discussed above.

\textit{(\textbf{ii}) Discrete-continuous sector:} $\mathcal{R}_{\left(
dis,con\right) }\left( C^{\ast }\right) $. \ \ Here the two abelian subgroup
representations have one discrete and one continuous spectrums either; $%
\mathcal{R}_{dis}\left( \mathbf{R}^{\ast }\right) \sim \mathbf{Z}$, and $%
\mathcal{R}_{con}\left( U\left( 1\right) \right) \sim \mathbf{R}$ and so,
\begin{equation}
\mathcal{R}_{\left( dis,con\right) }\left( C^{\ast }\right) \sim \mathbf{%
Z\times R.}
\end{equation}
\textit{or a continuous-discrete sector: }$\mathcal{R}_{\left(
con,dis\right) }\left( C^{\ast }\right) $ where now $\mathcal{R}_{con}\left(
\mathbf{R}^{\ast }\right) \sim \mathbf{R}$, \ but $\mathcal{R}_{dis}\left(
U\left( 1\right) \right) \sim \mathbf{Z}$; $\ \mathcal{R}_{\left(
con,dis\right) }\left( C^{\ast }\right) \sim \mathbf{R\times Z}$. \ \
\textit{(iii) }Finally\textit{\ the continuous-continuous sector }$\mathcal{R%
}_{\left( con,con\right) }\left( C^{\ast }\right) \sim \mathbf{R\times R}$
where both subgroups representations have continuous spectrums.\textbf{\ }We
will give more details on this fourth kind of representations once we finish
the remarks.

The second remark concerns the case of $C^{\ast r}$ toric symmetries
generated by the $U_{a}$ operators and the $V_{\tau _{a}}=V_{a}$
automorphisms, eq(3.15) extends, in the simplest situation, as $%
U_{a}V_{b}=\delta _{ab}\exp \left( -i\psi _{a}\tau _{a}\right) $ $V_{b}U_{a}$%
. More general extensions of eq(3.15) may be also written down; for more
details see \cite{a18} for similar realizations concerning discrete
symmetries. The last remark we give deals with the shift operator $V_{\tau }$%
; like for the $C^{\ast }$ group element $U$, the operator $V_{\tau }$ is an
element of the $C^{\ast }$ dual group, denoted as $\widetilde{C}^{\ast }$,
and acting on $C^{\ast }$\ as $VUV^{-1}=e^{i\psi \tau }U$. So the groups $%
C^{\ast }$\ and $\widetilde{C}^{\ast }$ do not commute; i.e \
\begin{equation}
C^{\ast }\widetilde{C}^{\ast }\neq \widetilde{C}^{\ast }C^{\ast }
\end{equation}
Later on we will consider the other case where two different $C^{\ast }$
factors of the toric group $C^{\ast r}$ do not commute as well. For the
moment we turn to complete our discussion by describing briefly the
continuous spectrums.

\textbf{b) Continuous Case}\newline
In this case the generator $T$ of $\mathcal{R}_{\left( con,con\right)
}\left( C^{\ast }\right) $\ has a continuous spectrum with a vector basis
state $\left\{ |z>,z\in \mathbf{C;\quad <}z\mathbf{^{\prime }|}z\mathbf{%
>=\delta }\left( z-z^{\prime }\right) \right\} $\ and acts diagonally as \ $%
<z|T|=z<z|$. The representation of the group element $U$ is $<z|U=\left(
\exp i\psi z\right) <z|$ which may also be put in the form $U=\int dz$ $\chi
\left( \psi ,z\right) $ $\pi \left( z\right) $ where the continuous $\chi
\left( \psi ,z\right) $\ character function is given by $\exp \left( i\psi
z\right) $ and where the $\pi \left( z\right) $'s, $\pi \left( z\right)
=|z><z|$; $\pi \left( z\right) \pi \left( z^{\prime }\right) =\mathbf{\delta
}\left( z-z^{\prime }\right) $ $\pi \left( z\right) $, are the\ projectors
on the $|z>$ states. Since the representation $\mathcal{R}_{\left(
con,con\right) }\left( C^{\ast }\right) $ is completely reducible, the $%
I_{id}$ identity operator may be decomposed as,
\begin{equation}
I_{id}=\int dz\text{ }\pi \left( z\right) .
\end{equation}
The shift operator by an $\epsilon $\ amount, denoted as $V\left( \epsilon
\right) $, acts on $\left\{ |z>,z\in \mathbf{C}\right\} $ as $<z|V\left(
\epsilon \right) =<z+\epsilon |$. It may be defined, by help of $\mathbf{a}%
^{+}\left( z,\epsilon \right) =|z><z+\epsilon |$ \ operators,\ as,
\begin{equation}
V\left( \epsilon \right) =\int dz\text{ }\mathbf{a}^{+}\left( z,\epsilon
\right) .
\end{equation}
These operators satisfy similar relations as in eqs(3.8-9) namely $UV=\exp
\left( -i\psi \tau \right) $ $VU$ and $U^{k}V^{l}=\left( \exp \left[ -i\psi
\tau \right] \right) ^{kl}V^{l}U^{k}$. This realization may also be defined
on the space of holomorphic functions $F\left( z\right) =\mathbf{<}z\mathbf{|%
}F\mathbf{>}$\ where
\begin{equation}
UFU^{-1}=\left( \exp i\psi z\right) \text{ }F,\quad V_{\epsilon
}FV_{\epsilon }^{-1}=\left( \exp \epsilon \partial _{z}\right) \text{ }F
\end{equation}

\paragraph{\textbf{2. NC }$C^{\ast }$\textbf{toric cycles}}

Here\ we forget about the $\widetilde{C}^{\ast }$ quantum symmetry and its $%
V_{a}$ generators and focus our attention on the $C^{\ast }$ toric
generators $U_{a}$ only. Of course a more general representation should
include also the $V_{a}$\ shift operators; it will be given later on; but as
far the $U_{a}$'s\ are concerned, one may also build representations where
the $r$ complex cycles of the $C^{\ast r}$ group are non commuting. This is
achieved by introducing torsion among the $C^{\ast }$ subgroups of the toric
symmetry by demanding the $T_{a}$\ generators to not commute; $\left[
T_{a},T_{b}\right] \neq 0$; say $\left[ T_{a},T_{b}\right] =im_{\left[ ab%
\right] }$. This means that given two toric actions $C^{\ast }$ and $C^{\ast
\prime }$, we have
\begin{equation}
C^{\ast }C^{\ast \prime }\neq C^{\ast \prime }C^{\ast }
\end{equation}
Here also we should distinguish between discrete and continuous spectrums.
In the particular case of a continuous spectrum, the $U_{a}=\lambda
_{a}^{T_{a}}=\exp i\psi _{a}T_{a}$, the algebra of NC $C^{\ast r}$ toric
cycles is defined as
\begin{equation}
U_{a}U_{b}=\Lambda _{ab}^{m_{\left[ ab\right] }}\text{ }U_{b}U_{a};\quad
\Lambda _{ab}=\exp \left( -i\psi _{a}\psi _{b}\right) ,\quad \left[
T_{a},T_{b}\right] =im_{\left[ ab\right] }I_{id}.
\end{equation}
The $r\times r$\ matrix $m_{ab}$ carries the $C^{\ast r}$ group torsion. A
possible realization of the generators $T_{a}$ on the space $\mathcal{F}$ of
holomorphic functions $f\left( y_{1},...,y_{r}\right) $ with $r$ arguments,
is that given by,
\begin{equation}
\left[ T_{a},f\left( y_{1},...,y_{r}\right) \right] =\left( \partial
_{a}-im_{ac}y_{c}\right) f.
\end{equation}
From this realization, one can check that the $T_{a}$\ generators satisfy
indeed the algebra $\left[ T_{a},T_{b}\right] =im_{\left[ ab\right] }I_{id}$%
. Note that eigen functions $f\left( y_{1},...,y_{r}\right) $ \ of $T_{a}$%
's\ with eigenvalues $k_{a}$\ are given by the holomorphic exponentiels $%
f_{\left( k_{1},...,k_{r}\right) }=\exp i\frac{m_{ab}k_{b}y_{a}^{2}}{2}$ ;
i.e
\begin{equation}
\left[ T_{a},f\right] =k_{b}f.
\end{equation}
These functions transform under finite transformations of the $C^{\ast r}$\
symmetry as
\begin{equation*}
U_{a}f_{\left( k_{1},...,k_{r}\right) }U_{a}^{-1}=\left( \exp i\psi
_{a}k_{a}\right) f_{\left( k_{1},...,k_{r}\right) }
\end{equation*}
Other representations of the $C^{\ast r}$ toric group with torsion may be
also written down; they are mainly obtained by complex extensions of known
results on non commutative real torii. However and as far as realizations of
NC toric group with torsions are concerned, one may introduce also the
quantum symmetries by allowing the \ $f\left( y_{1},...,y_{r}\right) $\ \
functions to depend on extra variables $z_{a}$ so that the new function is $%
F\left( z_{a};y_{a}\right) $ and the $\mathbf{C}^{\ast r}$ realization reads
as,
\begin{equation}
U_{a}FU_{a}^{-1}=\left( \exp i\psi _{a}\left( z_{a}+\partial
_{y_{a}}-im_{ac}y_{c}\right) \right) \text{ }F,\quad V_{b}FV_{b}^{-1}=\left(
\exp \epsilon _{bd}\partial _{z_{d}}\right) \text{ }F
\end{equation}
\ Having studied the main lines of NC $C^{\ast r}$ toric actions, we turn
now to solve the constraint eqs (3.4).

\subsubsection{Representations of the $Z_{i}$'s}

The constraint eqs(3.4) may be solved in different ways depending on whether
quantum symmetries of the $C^{\ast r}$\ actions are taken into account or
not. For the special example where only the $U_{a}$ generators of the $%
C^{\ast r}$ toric group are considered; then we can solve the $\theta _{ij}$
parameters by using the $m_{ab}$\ torsions. In the general case, one should
also be aware about the $\tau _{ab}$\ torsions between the $U_{a}$ and $%
V_{a} $ generators; i.e by using the relations $U_{a}V_{b}=\Omega
_{ab}^{\tau _{ab}}V_{b}U_{a}$. Here below, we shall give details for the
case where $\tau _{ab}=0$; but $m_{\left[ ab\right] }\neq 0$.

\textbf{Representation I}\newline
In this representation, the $Z_{i}$'s are realized in terms of the $%
U_{a}=\exp \left( i\psi _{a}T_{a}\right) $ as
\begin{equation}
Z_{i}=x_{i}\prod_{a=1}^{r}U_{a}^{q_{i}^{a}}=x_{i}\prod_{a=1}^{r}\exp \left(
iq_{i}^{a}\psi _{a}T_{a}\right) ,
\end{equation}
where $x_{i}$\ are complex moduli, which we shall interpret as just the
commutative coordinates of the toric manifold $\mathcal{V}_{d+1}$ containing
$\mathcal{M}_{d}$. Since the NC $C^{\ast r}$ toric group generators fulfill
relations such $U_{a}^{s}U_{b}=\Lambda ^{sm_{\left[ ab\right]
}}U_{b}U_{a}^{s}$ and \ $U_{a}U_{b}^{t}=\Lambda ^{tm_{\left[ ab\right]
}}U_{b}U_{a}^{t}$, it follows therefore that $%
U_{a}^{q_{i}^{a}}U_{b}^{q_{j}^{b}}=\Lambda ^{m_{\left[ ab\right]
}q_{i}^{a}q_{j}^{b}}$ $U_{b}^{q_{j}^{b}}U_{a}^{q_{i}^{a}}$ and consequently
\begin{equation}
\theta _{ij}=\prod_{a,b=1}^{r}\Lambda _{ab}^{m_{\left[ ab\right]
}q_{i}^{a}q_{j}^{b}}.
\end{equation}
Putting this solution back into eqs(3.4), one discovers that the constraint
eqs are indeed fulfilled because of the Calabi-Yau condition $%
\sum_{i=0}^{k}q_{i}^{a}=0$, but also due to the toric data encoded in$%
\sum_{i=0}^{k}q_{i}^{a}n_{\alpha i}=0$, as shown here below ,
\begin{eqnarray}
\prod_{i=0}^{k}Z_{i} &=&\left( \prod_{i=0}^{k}x_{i}\right)
\prod_{a=1}^{r}U_{a}^{\sum_{i=0}^{k}q_{i}^{a}}=\left(
\prod_{i=0}^{k}x_{i}\right) I_{id},  \notag \\
\prod_{i=0}^{k}Z_{i}^{n_{\alpha i}} &=&\left(
\prod_{i=0}^{k}x_{i}^{n_{\alpha i}}\right)
\prod_{a=1}^{r}U_{a}^{\sum_{i=0}^{k}q_{i}^{a}n_{\alpha i}}=\left(
\prod_{i=0}^{k}x_{i}\right) I_{id}.
\end{eqnarray}
Before going ahead note that such solutions may be extended by switching on
the $\tau _{ab}$ torsions. The result is $Z_{i}=x_{i}\prod_{a=1}^{r}\left(
U_{a}V_{a}\right) ^{q_{i}^{a}}$. Note moreover that according to the
spectrums of the group representation of the $U_{a}=\exp \left( i\psi
_{a}T_{a}\right) $ elements, the $Z_{i}$ operators may have also discrete or
continuous spectrums.

\textbf{Other Representations}\newline
The solutions we have given here above are still less general; they are
based on the two following remarkable identities; $\sum_{i=0}^{k}q_{i}^{a}=0$
and $\sum_{i=0}^{k}q_{i}^{a}n_{\alpha i}=0$. There are however other
relations similar to the previous ones and which can do the same job. These
relations are given by the identities $\sum_{i=0}^{k}Q_{i}^{aA}=%
\sum_{i=0}^{k}N_{i\alpha }^{a}=0$ defining the toric data eqs (2.17-18) of
the polygon of the Calabi-Yau manifold. Taking into account of these
identities, one may write down more general solutions extending eqs(3.28);
\begin{equation}
Z_{i}=x_{i}\prod_{a=1}^{r}\exp i\psi _{a}\left(
q_{i}^{a}+\sum_{A=1}^{d}\epsilon _{A}\text{ }Q_{i}^{aA}+\sum_{\alpha
}\epsilon ^{\alpha }\text{ }N_{i\alpha }^{a}\right) T_{a},
\end{equation}
where $\epsilon _{A}$\ and $\epsilon ^{\alpha }$\ are in general complex
numbers. The last two terms on the right hand of the above equation namely, $%
\left( \sum_{A=1}^{d}\epsilon _{A}\text{ }Q_{i}^{aA}+\sum_{\alpha }\epsilon
^{\alpha }\text{ }N_{i\alpha }^{a}\right) T_{a}$, constitute extra
contributions for the NC commutativity $\theta _{ij}$ parameters of the NC
Calabi-Yau manifold. They reflect the couplings between the $C^{\ast r}$
toric action and the toric data of the polygons $\Delta $ of the toric
manifold.

As a summary to this presentation, the $Z_{i}$ operator coordinates of the
NC Calabi-Yau manifold $\mathcal{M}_{d}^{nc}$ are generally realized,\ in
terms of the $U_{a}$ generators of the $C^{\ast r}$ toric group and their
underlying quantum symmetries generated by the $V_{a}$ shift operators, as
follows
\begin{equation}
Z_{i}=x_{i}\prod_{a=1}^{r}\left( U_{a}V_{a}\right) ^{\widetilde{q}_{i}^{a}}.
\end{equation}
Here $\widetilde{q}_{i}^{a}$, which are given by $\widetilde{q}%
_{i}^{a}=q_{i}^{a}+\sum_{A=1}^{d}\epsilon _{A}$ $Q_{i}^{aA}+\sum_{\alpha
}\epsilon ^{\alpha }$ $N_{i\alpha }^{a}$, are a kind of shifted Calabi-Yau
charges which satisfy the condition $\sum_{i=0}^{k}\widetilde{q}_{i}^{a}=0$
and their integrality follow by requiring $\epsilon _{A}$ and $\epsilon
^{\alpha }$ to be integer numbers. Using the relations
\begin{equation*}
U_{a}U_{b}=\Lambda _{ab}^{m_{\left[ ab\right] }}U_{b}U_{a}
\end{equation*}
and
\begin{equation*}
U_{a}V_{b}=\Omega _{ab}^{\tau _{ab}}V_{b}U_{a},
\end{equation*}
the non commutative $\theta _{ij}$\ parameters are expressed as
\begin{equation*}
\theta _{ij}=\prod_{a,b=1}^{r}\Lambda _{ab}^{J_{ij}^{ab}}\Omega
_{ab}^{K_{ij}^{ab}},
\end{equation*}
where now $J_{ij}^{ab}=m_{\left[ ab\right] }\widetilde{q}_{i}^{a}\widetilde{q%
}_{j}^{b}$ and $K_{ij}^{ab}\sim \tau _{\left[ ab\right] }\widetilde{q}%
_{i}^{a}\widetilde{q}_{j}^{b}$. By appropriate choices of $\Lambda _{ab}$, $%
\Omega _{ab}$\ , $m_{\left[ ab\right] }$ and $\tau _{\left[ ab\right] }$,
one recovers as special cases the\ representations involving discrete
torsions obtained in refs \cite{a15,a17}.

\subsection{Example}

To illustrate the previous analysis, we consider the NC extension of a
Calabi-Yau manifold with a conic singularity. This manifold is defined as a
hypersurface $\mathcal{M}_{2}$ embedded in the complex three dimension toric
variety $\mathcal{V}_{3}\subset \mathbf{C}^{6}\mathbf{/C}^{\ast 2}$ with a $%
\mathbf{C}^{\ast 2}$ toric actions defined as $x_{i}\longrightarrow
x_{i}\exp i\left( \psi _{a}q_{i}^{a}\right) $ and $q_{i}^{a}$ charges taken
as; $q_{i}^{1}=\left( 1,-1,1,-1,0,0\right) ,\quad q_{i}^{2}=\left(
1,0,-1,1,-1,0\right) $ and $p_{\alpha }=\left( 1,-1,1,-1\right) $. The $%
\mathbf{\nu }_{i}$ and $\mathbf{\nu }_{\alpha }^{\ast }$\ vertices of the
polygon $\Delta $ of the toric manifold and it dual $\nabla $, satisfying $%
\sum_{i=1}^{6}q_{i}^{a}\mathbf{\nu }_{i}=0$ and $\sum_{\alpha
=1}^{4}p_{\alpha }\mathbf{\nu }_{\alpha }^{\ast }=0$, are given the
following four dimensional vectors with integer entries $\nu _{i}^{A}=\left(
\nu _{i}^{1},\nu _{i}^{2},\nu _{i}^{3},\nu _{i}^{4}\right) $ and $\nu
_{iA}^{\ast }=\left( \nu _{i1}^{\ast },\nu _{i2}^{\ast },\nu _{i3}^{\ast
},\nu _{i4}^{\ast }\right) $ respectively,
\begin{equation}
\mathbf{\nu }_{i}=\left(
\begin{array}{cccc}
1 & 1 & 0 & 0 \\
1 & 2 & 1 & 0 \\
1 & 1 & 2 & 1 \\
1 & 0 & 1 & 1 \\
1 & 0 & -1 & 0 \\
1 & n_{2}+n & n_{2} & m
\end{array}
\right) ,\quad \mathbf{\nu }_{\alpha }^{\ast }=\left(
\begin{array}{cccc}
1 & 0 & 0 & 0 \\
1 & 1 & -1 & 0 \\
1 & 2 & -2 & 1 \\
1 & 1 & -1 & 1
\end{array}
\right) .
\end{equation}
The four $u_{\alpha }$ gauge invariants read as $u_{\alpha
}=\prod_{i=1}^{6}x_{i}^{n_{\alpha i}}$, $i=1,...,6$ \ with $n_{\alpha i}$\
integers given by,
\begin{equation}
n_{\alpha i}=\left(
\begin{array}{cccccc}
1 & 1 & 1 & 1 & 1 & 1 \\
2 & 2 & 0 & 0 & 2 & n+1 \\
3 & 3 & 0 & 0 & 3 & 2n+m+1 \\
2 & 2 & 1 & 1 & 2 & n+m+1
\end{array}
\right) .
\end{equation}
They satisfy the constraint eq $n_{0i}+n_{2i}=n_{1i}+n_{3i}$, showing in
turns that the complex three dimension toric manifold $\mathcal{V}_{3}$\ is
described by $u_{0}u_{2}=u_{1}u_{3}$ having a conic singularity at the
origin. The complex two dimension Calabi-Yau hypersurface embedded in $%
\mathcal{V}_{3}$ , $\sum_{\alpha }u_{\alpha }=0$, \ reads, in terms of the $%
x_{i}$ local coordinates, as
\begin{equation}
P\left( x_{1},...,x_{6}\right)
=a\prod_{i=1}^{6}x_{i}+x_{1}^{2}x_{2}^{2}x_{6}^{n+1}\left( x_{5}^{2}+b\text{
}x_{1}x_{2}x_{5}^{3}x_{6}^{n+m}+c\text{ }x_{3}x_{4}x_{5}^{2}x_{6}^{m}\right)
,
\end{equation}
where $n$ and $m$ are positive integers which may be fixed to some values.

The non commutative extension $\mathcal{M}_{2}^{nc}$ of this holomorphic
hypersurface is given by the NC algebra (3.2),\ generated by the $Z_{i}$\
generators satisfying eqs(3.28-31) with $\theta _{ij}=\Lambda ^{L_{ij}}$\
parameters given by eqs(29). For the special case where the $L_{ij}$
antisymmetric matrix is restricted to $L_{ij}=m\left(
q_{i}^{1}q_{j}^{2}-q_{j}^{1}q_{i}^{2}\right) $, with $L_{ij}=-L_{ji}$ and $%
L_{i6}=0$ and entries,
\begin{equation}
L_{ij}=m\left(
\begin{array}{ccccc}
0 & 1 & -2 & 2 & -1 \\
-1 & 0 & 1 & -1 & 1 \\
2 & -1 & 0 & 0 & -1 \\
-2 & 1 & 0 & 0 & 1 \\
1 & -1 & 1 & -1 & 0
\end{array}
\right) .
\end{equation}
The NC complex surface $\mathcal{M}_{2}^{nc}$ is then given by a one
parameter algebra generated by following relations,
\begin{eqnarray}
Z_{1}Z_{2} &=&\Lambda ^{m}\text{ }Z_{2}Z_{1},\quad Z_{1}Z_{3}=\Lambda ^{-2m}%
\text{ }Z_{3}Z_{1},\qquad Z_{1}Z_{4}=\Lambda ^{2m}\text{ }Z_{4}Z_{1},  \notag
\\
Z_{1}Z_{5} &=&\Lambda ^{-m}\text{ }Z_{5}Z_{1},\qquad Z_{2}Z_{3}=\Lambda ^{m}%
\text{ }Z_{3}Z_{2},\qquad Z_{2}Z_{4}=\Lambda ^{-m}\text{ }Z_{4}Z_{2},  \notag
\\
Z_{2}Z_{5} &=&\Lambda ^{m}\text{ }Z_{5}Z_{2},\qquad Z_{3}Z_{4}=\text{ }%
Z_{4}Z_{3},\qquad Z_{3}Z_{5}=\Lambda ^{-m}\text{ }Z_{5}Z_{3}, \\
Z_{4}Z_{5} &=&\Lambda ^{m}\text{ }Z_{5}Z_{4},\qquad Z_{i}Z_{6}=Z_{6}Z_{i},
\notag
\end{eqnarray}
where $\Lambda ^{m}$\ is given by $\Lambda ^{m}=\exp \left( -im\psi _{1}\psi
_{2}\right) $. Since $\psi _{a}=\rho _{a}-i\alpha _{a}$; it follows that $%
m\psi _{1}\psi _{2}=m\left( \rho _{1}\rho _{2}-\alpha _{1}\alpha _{2}\right)
-im\left( \alpha _{1}\rho _{2}+\alpha _{2}\rho _{1}\right) $ which we set as
$\Lambda ^{m}=\exp \left( \kappa +i\phi \right) $\ for simplicity. This is a
complex parameter enclosing various special situations corresponding to: (1)
\ \textit{Hyperbolic representation} described by $\left( \kappa ,\phi
\right) \equiv \left( \kappa ,0\right) $; it corresponds to torsions induced
by $\mathbf{R}^{\ast 2}\otimes U\left( 1\right) ^{2}$ subgroup of the $%
C^{\ast 2}$ toric group. (2) \textit{Periodic representations} corresponding
to $\left( \kappa ,\phi \right) \equiv \left( 0,\phi +2\pi \right) $ where $%
\left| \Lambda ^{m}\right| =1$. It is associated with a NC $\mathbf{R}^{\ast
}\ast U\left( 1\right) $ toric actions of the $C^{\ast 2}$ group. (3)
\textit{Discrete periodic representations} $\left( \kappa ,\phi \right)
\equiv \left( 0,N\phi +2\pi \right) $ with $\left| \Lambda ^{m}\right| =1$
but moreover $\left( \Lambda ^{m}\right) ^{N}=1$. This last case is
naturally a subspace of the periodic representation and it is precisely the
kind of situation that happen in the building of NC manifolds with discrete
torsion. It is associated with the $\mathbf{\ Z}_{N}$ subgroup of $U\left(
1\right) $.

\section{Fractional Branes}

The NC type $IIA$ realization we have studied so far concerns regular points
of the algebra, that is non singular ones of the $\mathbf{C}^{\ast r}$ toric
group. In this section, we want to complete this analysis by considering the
representations at $C^{\ast r}$ group fixed points where it is expected to
get fractional $D$ branes. To do so, we shall first classify the various $%
\mathcal{M}_{d}$ subsets $\mathcal{S}_{(\mu )}$ of stable points under $%
\mathbf{C}^{\ast r}$; then we give the quiver diagrams extending those of
Berenstein and Leigh obtained for the case of orbifolds with discrete
torsion. The method is quite similar to that of \cite{a24}.

\subsection{Fixed points of the $C^{\ast r}$\ toric actions}

\qquad To fix the idea, consider that class of Calabi-Yau manifolds $%
\mathcal{M}_{d}$ described by hypersurfaces $P_{d}\left[ x_{1},...,x_{k+1}%
\right] $ eq(2.16) embedded in $\mathcal{V}_{d+1}$, with toric data $\left\{
q_{i}^{a},\mathbf{\nu }_{i};p_{\alpha }^{I},\mathbf{\nu }_{\alpha }^{\ast
}\right\} $ satisfying eqs(2.4). The local holomorphic coordinates $\left\{
x_{i}\in \mathbf{C}^{k+1};\quad 0\leq i\leq k\right\} $ are not all of them
independent as they are related by the $\mathbf{C}^{\ast r}$ gauge
transformations $U_{a}:x_{i}\longrightarrow
U_{a}x_{i}U_{a}^{-1}=x_{i}\lambda ^{q_{i}^{a}}$, with $%
\sum_{i=0}^{k}q_{i}^{a}=0$.\ Fixed points of the $\mathbf{C}^{\ast r}$ gauge
transformations are given by the solutions of the constraint eq
\begin{equation}
x_{i}=U_{a}x_{i}U_{a}^{-1}=x_{i}\lambda _{a}^{q_{i}^{a}}.
\end{equation}
From this relation, one sees that its solutions depend on the values of $%
q_{i}^{a}$; the $x_{i}$'s should be zero unless $q_{i}^{a}=0$. Fixed points
of $\mathbf{C}^{\ast r}$ toric actions are then given by the $\mathcal{S}$
subspace of $\mathbf{C}^{k+1}$ whose $x_{i}$ local coordinates are $\mathbf{C%
}^{\ast r}$\ gauge invariants. To get a more insight into this subspace it
is interesting to note that as $\mathbf{C}^{k+1}\mathbf{/C}^{\ast r}=\left(
\mathbf{C}^{k+1}\mathbf{/C}^{\ast }\right) \mathbf{/C}^{\ast r-1}$ ; it is
useful to introduce the $\mathcal{S}_{(a)}=\left\{ x_{i};q_{i}^{a}=0;\qquad
0\leq i\leq k\right\} $\ subspaces that are invariant under the $a-th$
factor of the $\mathbf{C}^{\ast r}$ group. So the manifold $\mathcal{S}$
stable under $\mathbf{C}^{\ast r}$ is given by the intersection of the
various $\mathcal{S}_{(a)}$'s,

\begin{equation}
\mathcal{S}=\cap _{a=1}^{r}\mathcal{S}_{(a)}
\end{equation}
If we suppose that $\left\{ x_{i_{0}};...;x_{i_{k_{0}-1}}\right\} $ those
local coordinates that have non zero $q_{i}^{a}$ charges and $\left\{
x_{i_{k_{0}-1}};...;x_{i_{k}}\right\} $\ the coordinates that are fixed
under $\mathbf{C}^{\ast r}$\ actions; then the manifold $\mathcal{S}$ is
given by,
\begin{equation}
\mathcal{S}=\left\{ \left( 0,...,0,x_{k_{0}},...,x_{k}\right) \right\}
\subset \mathcal{V}_{d+1}\subset \mathbf{C}^{k+1}
\end{equation}
To get the representation of the $Z_{i}$ variable operators on the $\mathcal{%
S}$\ space, let us first consider what happens on its neighboring space $%
\mathcal{S}^{\epsilon }=\left\{ \left( \epsilon ,...,\epsilon
,x_{k_{0}}+\epsilon ,...,x_{k}+\epsilon \right) \right\} $, where we have
taken $\epsilon _{0}=...=\epsilon _{k}=\epsilon $ \ and where $\epsilon $\
is as small as possible. Using the hypothesis $%
q_{i_{k_{0}}}^{a}=...=q_{i_{k}}^{a}=0$ \ we have made and replacing the $%
x_{i}$ moduli by their expression on $\mathcal{S}^{\epsilon }$, then putting
in the realization eqs(3.28), we get the following result,

\begin{eqnarray}
Z_{i_{j}} &=&\epsilon \prod_{a=1}^{r}U_{a}^{q_{i_{j}}^{a}};\quad 0\leq j\leq
k_{0}-1, \\
Z_{i_{j}} &=&\left( x_{i_{j}}+\epsilon \right) I_{id};\quad k_{0}\leq j\leq
k.
\end{eqnarray}
The representation of the $Z_{i}$'s\ on the space $\mathcal{S}_{0}$\ is then
obtained by taking the limit $\epsilon \longrightarrow 0$.\ As such non
trivial operators are given by $Z_{i_{j}}\sim x_{i_{j}}I_{id};\quad
k_{0}\leq j\leq k$; they are proportional to the identity $I_{id}$ operator
of\ group representation $\mathcal{R}\left( C^{\ast r}\right) $. This an
important point since the $Z_{i}$\ operators are reducible into an infinite
component sum as shown here below,\footnote{%
The sums involved in the decomposition of the identity are either discrete
series, integrals or both of them depending on whether the group
representation $\mathcal{R}\left( C^{\ast r}\right) $ spectrum is discrete,
continuous or with discrete and continuous sectors. Therefore we have either
$I_{id}=\sum_{\mathbf{n}}\pi _{\mathbf{n}}$, $\ \mathbf{n=}\left(
n_{1},...,n_{r}\right) $; $\ I_{id}=\int d\mathbf{\sigma }\pi \left( \mathbf{%
\sigma }\right) $, $\mathbf{\sigma =}\left( \sigma _{1},...,\sigma
_{r}\right) $; or again $\sum_{\mathbf{m}}\int d\mathbf{\zeta }\pi _{\mathbf{%
m}}\left( \mathbf{\zeta }\right) $; Here $\ \pi _{\mathbf{m}}$, $\ \mathbf{m=%
}\left( n_{i_{1}},...,n_{i_{r_{0}}}\right) $ \ and $\pi \left( \mathbf{\zeta
}\right) $, \ $\mathbf{\zeta =}\left( \zeta _{i_{r_{0}+1}},...,\zeta
_{i_{r}}\right) $; \ and are the $C^{\ast r}$ representation projectors
considered in section 3 .}
\begin{equation}
Z_{i}=\sum_{\mathbf{n}}\text{ }Z_{i}^{\left( \mathbf{n}\right) };\quad
Z_{i}^{\left( \mathbf{n}\right) }=x_{i}\pi _{\mathbf{n}}.
\end{equation}
The above decomposition of the $Z_{i}$'s on $\mathcal{S}$ has a nice
interpretation in the $D$ brane language. Thinking of the $x_{i}$ variables
as the coordinates of a $D$ $p$ brane, ( $p=2d$), wrapping the Calabi-Yau
manifold $M_{d}$, it follows that, due torsion of the $C^{\ast r}$ toric
group, the $D$ $p$ brane at the singular points fractionate in the same
manner as in the analysis of Berenestein and Leigh for the case of orbifolds
with discrete isometries. In addition to the results of \cite{a24}, which
apply as well for the present study, there is a novelty here due to the
dimension of the completely reducible $\mathcal{R}\left( C^{\ast r}\right) $
group representation. There are infinitely many values for the $C^{\ast }$
characters and so an infinite number of fractional $D\left( p-2k_{0}\right) $
branes wrapping $\mathcal{S}$.

\subsection{Quiver Diagrams}

Like for the case of Calabi-Yau orbifolds with discrete symmetries, one can
here also describe the varieties of fractional $D$ branes by generalizing
the Berenstein and Leigh quiver diagrams for $\mathcal{M}_{d}^{nc}$ at the
fixed points of the $\mathbf{C}^{\ast r}$\ toric actions. One of the basic
ingredients in getting these graphs is the identification of the projectors
of the $\mathbf{C}^{\ast r}$ toric action\ group and the step operators $%
\mathbf{a}^{\pm }$ acting as shift operators on the basis states of the
group representation space. Since these actions are given by a kind of
complexification of $U\left( 1\right) ^{r}$ and as each $\mathbf{C}^{\ast }$
group factor has completely reducible representations with four possible
sectors, it is interesting to treat separately these different cases. The
various sectors for each $\mathbf{C}^{\ast }$ subsymmetry factor are as
follows; (i) $\left( dis,dis\right) $ discrete-discrete sector where the $%
C^{\ast }$ characters $\chi _{n}\left( \psi \right) $ are given $\chi
_{n}\left( \psi \right) =\exp i\psi n$; $n\in \mathbf{Z+}i\mathbf{Z,}$ $\psi
\in \mathbf{C}$; (ii) $\left( dis,con\right) $ discrete-continuous and \ $%
\left( con,dis\right) $ \ continuous-discrete sectors \ and finally (iii) \ $%
\left( con,con\right) $ continuous-continuous sector with characters $\chi
\left( p,\psi \right) $ given by $\chi \left( p,\psi \right) =\exp ip\psi $;
$p\in \mathbf{C}$.

Recall first of all that, due to torsion of the $C^{\ast r}$ toric
symmetries, the algebraic structure of the $D$ $p$ branes wrapping the
compact manifold $\mathcal{M}_{d}$ change. Brane points $\{x_{i}\}$ of
commutative type $IIA$ geometry become, in presence of torsion, fibers based
on $\{x_{i}\}$. These fibers are valued in the algebra of the group
representation $\mathcal{R}\left( C^{\ast r}\right) $ and may be given a
simple graph description on fixed spaces. While points $x_{i}.1$ in the
commutative type $IIA$ geometry are essentially numbers, the $Z_{i}$
coordinate operators can be thought of, in the case of a discrete spectrum
of $\mathbf{C}^{\ast }$, as
\begin{equation}
x_{i}.1=\ \rightarrow \ \ Z_{i}=\left( Z_{i}\right) _{mn}\ U^{m}V^{n},
\end{equation}
where $U$ and $V$, $UV=e^{-i\psi }VU$, \ are the generators of the $C^{\ast
} $ toric group

Extending the results of [24], one can draw graphs for fractional $D$
branes. Due to the decomposition of $I_{id}$ eqs(3.13) and (3.20), we
associate to each $D$ $p$ brane coordinate a quiver diagram mainly given by
the product of ( discrete or continuous )\ $S^{1}$\ circles. For the
simplest case $r=1$ and $C^{\ast }$ discrete representations, the quiver
diagram is built as follows:

(1) To each $\pi _{n}=|n><n|$ projector it is associated a vertex point on a
discrete $S^{1}$ circle. As there is an infinite number of points that one
should put on $S^{1}$, all happens as if the quiver diagram is given by the $%
\mathbf{Z+}i\mathbf{Z}$ \ lattice plus a extra point at infinity.

(2) The $a_{n}^{\pm }=|n\pm 1><n|$ shift operators are associated with the
oriented links joining adjacent vertices, ( vertex $\left( n-1\right) $ to
the vertex $\left( n\right) $ for $a_{n}^{-}$\ and vertex $\left( n\right) $
to the vertex $\left( n+1\right) $ for $a_{n}^{+}$\ ), of quiver diagram.
They act as automorphisms exchanging the $\mathbf{C}^{\ast }$\ characters.

Moreover, as non zero $D$ $p$ brane coordinates at the singularities are of
the form $Z_{i}\sim \sum_{n}$ $Z_{i}^{\left( n\right) }$, it follows that $D$
$p$ branes on $\mathcal{S}$ sub-manifolds fractionate into an infinite set
of fractional $D2s$ branes coordinated by $Z_{i}^{\left( n\right) }$. This
is a remarkable feature which looks like the inverse process of tachyon
condensation mechanism \`{a} la GMS \cite{a8} where for instance a $D$ $%
p_{1} $ brane on a NC Moyal plane decomposes into an infinite set of $D$ $%
\left( p_{1}-2\right) $ branes. For details see studies on branes and
Noncommutative Solitons \cite{a8},\cite{a30}-\cite{a32} in particular $D25$
branes decaying into an infinite $D23$ ones. In the case of a continuous
spectrum, the corresponding quiver diagram is given by cross products of
circles.

\section{Conclusion}

Using the algebraic geometry approach of Berenstein and Leigh, we have
studied\ the type $IIA$ geometry of non commutative Calabi-Yau manifolds
embedded in non commutative toric varieties $\mathcal{V}$. Actually this
study completes partial results of works in the literature on NC Calabi-Yau
manifolds and too particularly orbifods of Calabi-Yau homogeneous
hypersurfaces with discrete torsion. Our construction has also the
particularity of going beyond the idea of Berenstein and Leigh by
introducing non commutative toric actions $\left( \mathbf{C}^{\ast r}\right)
^{nc}$ involving NC complex torii generalizing the Connes et \textit{al}
ones used in the study of matrix model compactification. From field
theoretic point of view, our way of doing may be thought of as a step for
approaching non commutative extension of supersymmetric gauged linear sigma
models and their Landau Ginzburg mirrors.

The results established in this paper concerns non commutative extension of
the class of Calabi-Yau manifolds $\mathcal{M}_{d}$ \ embedded in toric
varieties $\mathcal{V}_{d+1}$ with $\mathbf{C}^{\ast r}$ toric actions
endowed by asymmetries. The latters are completely specified by the toric
data
\begin{equation}
\left\{ q_{i}^{a};\nu _{i}^{A};p_{\alpha }^{I};\nu _{\alpha A}^{\ast };\quad
0\leq i\leq k;\quad 1\leq a\leq r;\quad 1\leq A\leq d+1\right\} ,
\end{equation}
with Calabi-Yau condition $\sum_{i=0}^{k}q_{i}^{a}=0$, the toric geometry
relations $\sum_{i=0}^{k}q_{i}^{a}\nu _{i}^{A}=0$\ and $\sum_{\alpha
=0}^{d+1+r^{\ast }}p_{\alpha }^{I}\nu _{\alpha A}^{\ast }=0;$ $%
I=1,...,r^{\ast }$. These eqs\ define the toric polygons of the variety $%
\mathcal{V}_{d+1}$. Non commutative structure is carried either by quantum
symmetries described by inner automorphisms of $\mathbf{C}^{\ast r}$ or
again by considering NC complex cycles within the toric group in the same
manner as one does in the Connes et \textit{al} approach of toroidal
compactification of matrix model of M theory. In our present case non
commutative structure is indeed solved in terms of asymmetries of the $%
\mathbf{C}^{\ast r}$ toric actions and the toric data of the underlying $%
\mathcal{V}_{d+1}$. This result extends partial ones on NC geometries using
discrete torsion of isometries of orbifolds of Calabi-Yau homogeneous
hypersurfaces which are recovered as particular cases. Among our main
results, we quote the two following:

(1) A class of complex $d$ dimension\ NC Calabi-Yau manifolds $\mathcal{M}%
_{d}^{nc}$ is naturally described in the language of toric geometry. They
are given by subalgebras of NC toric varieties $\mathcal{V}_{d+1}^{nc}\sim
\mathbf{C}_{\theta }^{k+1}\mathbf{/C}_{m,\tau }^{\ast r}$ where the NC
matrix parameter $\theta _{ij}$ is induced by asymmetries of the $\mathbf{C}%
^{\ast r}$ toric group. In this picture discrete groups $\Gamma $ may be
also included by taking into account the discrete symmetries of the toric
variety generally described by $\prod_{\alpha }u_{\alpha }^{p_{\alpha
}^{I}}=1$. For both kinds of NC toric varieties $\mathcal{V}_{d+1}^{nc}\sim
\mathbf{C}_{\theta }^{k+1}\mathbf{/C}_{m,\tau }^{\ast r}$ and $\mathcal{V}%
_{d+1}^{nc}\sim \mathbf{C}_{\Theta }^{k+1}\mathbf{/C}_{m,\tau }^{\ast
r}\times \Gamma $, where in addition to $\theta _{ij}$, the $\Theta _{ij}$\
parameters have extra contributions coming from discrete torsion of $\Gamma $%
, we have
\begin{equation}
\mathcal{M}_{d}^{nc}\subset \mathcal{V}_{d+1}^{nc};\quad \mathcal{M}_{d}=%
\mathcal{Z}\left( \mathcal{M}_{d}^{nc}\right) \subset \mathcal{V}_{d+1}=%
\mathcal{Z}\left( \mathcal{V}_{d+1}^{nc}\right)
\end{equation}
Results obtained in this way covers as special cases those derived by
following the method used in \cite{a15},\cite{a17},\cite{a24},\cite{a18} and
where no reference to the toric data are made; see also \cite{a34}.

(2) As far Calabi-Yau manifolds embedded in toric varieties are concerned,
we have shown that the $\theta _{ij}$ parameters of the non commutative
structure have contributions involving the toric data of polygons and, in
addition to the Calabi-Yau condition $\sum_{i=0}^{k}q_{i}^{a}=0$, it uses as
well the relations $\sum_{i=0}^{k}q_{i}^{a}\nu _{i}^{A}=0$ \ for the solving
of the constraint eqs. Note in passing that in the analysis of section 3, we
have considered the special projections of eq $\sum_{i=0}^{k}q_{i}^{a}\nu
_{i}^{A}=0$ on $v_{j}^{A}$ and $v_{jA}^{\ast }$\ namely; $%
\sum_{i=0}^{k}q_{i}^{a}\nu _{i}^{A}\cdot v_{j}^{A}=0$ \ and $%
\sum_{i=0}^{k}q_{i}^{a}\nu _{i}^{A}\cdot v_{jA}^{\ast }=0$. In general one
may also use other projections by generic vectors $\mathbf{u}_{A}$ of the $%
\mathbf{Z}^{d+1}$ lattice. An other remarkable feature of the $C^{\ast r}$
toric group is that at its fixed points we have an infinite set of
fractional $D$ branes instead of a finite one as it the case of NC orbifolds
by $\mathbf{Z}_{N}^{n}$ groups as shown in \cite{a15,a17}. This special
feature is similar to the tachyon condensation picture of string field
theory \cite{a8},\cite{a33}.

To do so, we have first studied the type $IIA$ geometry of complex $d$
dimension Calabi-Yau manifolds using toric geometry methods. Then we have
given the constraint eqs defining their NC geometry extensions by using the
Berenstein et \textit{al }method and second by considering embedding in NC\
toric varieties. To work out the regular solutions of the constraint eqs, we
have developed two realizations of the NC $C^{\ast r}$ toric group; one
involving quantum symmetries generated by shift operators $V_{a}$ on the
states of the $C^{\ast r}$ group representation and the other using the
torsions between the generators $U_{a}$ of the toric group. Next we have
given different classes of solutions depending the nature of torsions of
the\ $C^{\ast r}$ toric symmetries. As singular points of the toric actions
are completely characterized by the $q_{i}^{a}$ charges, we have studied
also the singular representations of the constraint eqs and analyzed
fractional $D$ branes at singularities. Since the representations of the
abelian $C^{\ast r}$ group are completely reducible, we have fractional $D$
branes at the singularities; but with the remarkable feature that now there
is an infinite set of them. This property follows naturally from the fact
that identity $I_{id}$ of the representation is decomposable into infinite
sums over the $\pi _{n}$ and $\pi \left( \sigma \right) $\ projectors namely
$I_{id}=\sum_{n}\pi _{n}$ for discrete spectrums and $I_{id}=\int d\sigma $ $%
\pi \left( \sigma \right) $ for continuous ones; see also footnote 4.
Actually this is a special feature of the non commutative structure induced
by torsions of continuous $C^{\ast r}$ groups; it is related to the
condensation phenomenon \`{a} la GMS considered few years ago in \cite{a8}
and subsequent works. As a perspective, it would be interesting to analyze
the properties of non commutative type $IIB$ geometry dual to type $IIA$ \
NC geometry considered in this paper; then explore the features of mirror
symmetry in the case NC Calabi-Yau manifolds. This study will be developed
in the second part of this work \cite{a35}.

\begin{acknowledgement}
We thank Dr Adil Belhaj for discussions and earlier collaborations on these
issues. We also thank the program Protars III/CNR/2003 , Rabat, for their support.\newpage
\end{acknowledgement}

\end{document}